\documentclass[oldversion]{aa}  
\usepackage{graphicx}
\usepackage{txfonts}
\usepackage{natbib}
\usepackage{epstopdf}
\bibpunct{(}{)}{;}{a}{}{,} 
\begin{document}
   \title{Dynamical properties of AMAZE and LSD galaxies from gas kinematics and the Tully-Fisher relation at $z\sim3$ \thanks{Based on observations collected with European Southern Observatory/Very Large Telescope (ESO/VLT) (proposals 075.A-0300, 076.A-0711 and 178.B-0838), with the Italian TNG, operated by FGG (INAF) at the Spanish Observatorio del Roque de los Muchachos, and with the {\it Spitzer Space Telescope}, operated by JPL (Caltech) under a contract with NASA. }}

   \titlerunning{Gas kinematics of AMAZE and LSD galaxies at $z\sim3$}


   \author{A. Gnerucci\inst{1}, A. Marconi\inst{1}, G. Cresci\inst{2}, R. Maiolino\inst{3}, F. Mannucci\inst{2}, F. Calura\inst{4}, A. Cimatti\inst{6}, F. Cocchia\inst{3}, A. Grazian\inst{3}, F. Matteucci\inst{5}, T. Nagao\inst{7}, L. Pozzetti\inst{8}, P. Troncoso\inst{3}
          }

   \authorrunning{Gnerucci et al.}

   \offprints{gnerucci@arcetri.astro.it}

   \institute{Dipartimento di Fisica e Astronomia, Universit\`a degli Studi di Firenze, Largo E. Fermi 2, 50125 Firenze, Italy\\
              \email{gnerucci@arcetri.astro.it}
              \and
              INAF-Osservatorio Astrofisico di Arcetri, Largo E. Fermi 5, 50125 Firenze, Italy\\
              \email{gcresci,filippo@arcetri.astro.it}
              \and
              INAF-Osservatorio Astronomico di Roma, via di Frascati 33, Monte Porzio Catone, Italy\\
              \email{maiolino@oa-roma.inaf.it}
              \and
              Jeremiah Horrocks Institute for Astrophysics and Supercomputing, University of Central Lancashire, Preston PR1 2HE, UK
              \and
              Dipartimento di Astronomia, Universit\`a di Trieste, via Tiepolo 11, 34131 Trieste, Italy
              \and
              Dipartimento di Astronomia, Universita` di Bologna, via Ranzani 1, 40127, Bologna, Italy
              \and
              Research Center for Space and Cosmic Evolution, Ehime University, 2-5 Bunkyo-cho, Matsuyama 790-8577, Japan
              \and
              INAF - Osservatorio Astronomico di Bologna, Via Ranzani 1, 40127 Bologna, Italy
             }

   \date{Received ; accepted}


 \abstract{
 
   {We present a SINFONI integral field kinematical study of 33  galaxies at $z\sim3$ from  the
   AMAZE and LSD projects which are aimed at studying metallicity and dynamics of high-redshift
   galaxies. The number of galaxies analyzed in this paper constitutes a significant
   improvement compared to existing data in the literature and this is the first time that
   a dynamical analysis
   is obtained for a relatively large sample of galaxies at  $z\sim3$.
   11 galaxies show ordered rotational motions ($\sim 30\%$ of the sample),
   in these cases we estimate
   dynamical masses by modeling the gas kinematics  with rotating disks and exponential mass
   distributions. We find dynamical masses in the range $2\times10^9 M_{\sun}-2\times 10^{11}
   M_{\sun}$ with a mean value of $\sim 2\times10^{10} M_{\sun}$. By comparing observed gas
   velocity dispersion with that expected from models, we find that most rotating
   objects are dynamically ``hot'', with intrinsic velocity dispersions of
   the order of $\sim 90$ km s$^{-1}$. The median value of the ratio between the maximum disk rotational velocity and the intrinsic velocity dispersion for the rotating objects is $1.6$, much lower than observed in local galaxies value ($\sim10$) and slightly lower than the $z\sim2$ value ($2-4$). Finally we use the maximum rotational velocity from our modeling to build a baryonic Tully-Fisher relation at z$\sim$3. Our measurements
   indicate that $z\sim3$ galaxies have lower stellar masses (by a factor of ten on average)
   compared to local galaxies with the same dynamical mass.
   However, the large observed scatter  suggests that
   the Tully-Fisher relation is not yet ``in place'' at these early cosmic ages, possibly due
   to the young age of galaxies. A smaller dispersion of the Tuly-Fisher relation is obtained by taking into account the velocity dispersion with the use of the $S_{0.5}$ indicator, suggesting that turbulent motions might have an important dynamical role.
}


   }
   \maketitle


\section{Introduction}\label{s1}

The dynamical properties of galaxies play
a fundamental role in the context of galaxy formation and evolution.
The observed dynamics of galaxies represent a test to theoretical models and are the most direct way to probe the content of dark matter.
In particular, the dynamical mass of a galaxy inferred from its rotation curve is the most direct way to constrain mass and angular momentum of dark matter haloes, 
which can be compared with the predictions of cosmological cold dark matter models of hierarchical structure formation.
In such models (\citealt{Blumenthal:1984}; \citealt{Davis:1985}; \citealt{Springel:2006}, \citealt{Mo:1998}) mergers are believed to play an important role for galaxy formation and evolution. However, the observational evidence for the existence of rotating disks with high star formation rates at $z\sim 2$
suggests that smooth accretion of pristine gas is also an important mechanism that drives
star formation and mass assembly at high redshift (\citealt{Epinat:2009}; \citealt{Wright:2009}; \citealt{Forster-Schreiber:2009}; \citealt{Cresci:2009}). 

In recent years, our observational knowledge in this field has increased enormously.
Many dynamical studies have been performed on extended samples of $z\sim1 -2.5$ objects (\citealt{Genzel:2006}; \citealt{Genzel:2008}; \citealt{Forster-Schreiber:2006}; \citealt{Forster-Schreiber:2009};
\citealt{Cresci:2009}; \citealt{Erb:2006}; \citealt{Law:2007}; \citealt{Law:2009}; \citealt{Epinat:2009}; \citealt{Wright:2007}; \citealt{Wright:2009}). However, little is known on the
dynamics of galaxies at $z\gtrsim2.5$, where only a few particular objects have been 
investigated
(\citealt{Nesvadba:2006}; \citealt{Nesvadba:2007}; \citealt{Nesvadba:2008};  \citealt{Jones:2010}; \citealt{Law:2009}; \citealt{Lemoine-Busserolle:2009}; \citealt{Swinbank:2007}; \citealt{Swinbank:2009}).

The redshift range $z\sim 3-4$ is particularly important to study since it is before the peak of the cosmic star formation rate (see, for example, \citealt{Dickinson:2003}; \citealt{Rudnick:2006}; \citealt{Hopkins:2006}; \citealt{Mannucci:2007}),
when only a small fraction ($\sim15\%$, \citealt{Pozzetti:2007}) of the stellar mass in present-day galaxies has been assembled.
It is also the redshift range when the most massive early-type galaxies are expected to form
(see, for example, \citealt{Saracco:2003}). Moreover, the number of galaxy mergers  is much larger than at later times (\citealt{Conselice:2007}; \citealt{Stewart:2008}). As a consequence the predictions of different models tend to diverge significantly at $z\gtrsim2.5$. 

This paper is based on two projects focused on studying metallicity and dynamics of  high-redshift galaxies: AMAZE (Assessing the Mass-Abundance redshift Evolution) (\citealt{Maiolino:2008}, \citealt{Maiolino:2008a}) and LSD (Lyman- break galaxies Stellar populations and Dynamics) (\citealt{Mannucci:2008}, \citealt{Mannucci:2009}). Both projects use integral field spectroscopy of samples of $z\sim3-4$ galaxies in order to derive their chemical and dynamical properties.
In both projects we make use of  data obtained with the Spectrograph for Integral Field Observations in the Near Infrared (SINFONI) at the Very Large Telescope (VLT) of the European Southern Observatory (ESO).
Integral field spectroscopy has proven to be a powerful tool to study galaxy dynamics as it provides two dimensional velocity maps, without the restrictions of longslit studies, also plagued by unavoidable light losses. 

The AMAZE sample consists of $\sim30$ Lyman Break Galaxies in the redshift range
$\rm 3<z<4.8$ (most of which at z$\sim$3.3), with deep Spitzer/IRAC photometry ($3.6-8\mu m$), an important piece of information to derive reliable stellar masses. These galaxies were observed with SINFONI in seeing-limited mode. The LSD sample is a representative, albeit small, sample of 10 LBGs at z$\sim$3 with available Spitzer and HST imaging from the \cite{Steidel:2003} catalogue. For LSD SINFONI observations were performed with the aid of adaptive optics in order to improve spatial resolution since this project was aimed to obtain spatially-resolved spectra for measuring kinematics and gradients in emission lines. The
global sample is listed in Tab.\ref{tab1}.

Lyman Break Galaxies are objects selected based on the Ly-break technique and
based on their UV rest-frame blue color. As a consequence these samples are biased both
against dust reddeneded systems and against aged stellar populations, which are characterized
by redder colors. Therefore, the conclusions inferred from our results apply only to
a sub-population (about half) of galaxies at z$\sim$3 (\citealt{Reddy:2005}, \citealt{van-Dokkum:2006}, \citealt{Grazian:2007}).

\begin{table*}
  \caption[!ht]{Galaxy properties and observation setups.}
  \label{tab1}
  \centering
  \begin{tabular}{l c c c c c c c}
    \hline
    \hline
    \noalign{\smallskip}
     Object & sample & R.A. $^{(1)}$ & Dec. $^{(1)}$ & z / scale(kpc/\arcsec) $^{(2)}$& Texp(min) & pixel scale(\arcsec) & line used $^{(3)}$\\
    \hline
    \hline
    \noalign{\smallskip}
    SSA22a-M38 &  AMAZE & 22:17:17.7  & +00:19:00.7 & $3.294/7.48$ & 400 & $0.125\times0.25$ & [OIII]\\
    SSA22a-C16 & AMAZE & 22:17:32.0 & +00:13:16.1 & $3.068/7.65$ & 350 & $0.125\times0.25$ & [OIII]\\
    CDFS-2528 & AMAZE & 03:32:45.5 & -27:53:33.3 & $3.688/7.18$  & 350 & $0.125\times0.25$ & [OIII] \\
    SSA22a-D17 & AMAZE & 22:17:18.9 & +00:18:16.8 & $3.087/7.64$ & 250 & $0.125\times0.25$ & [OIII]\\
    CDFA-C9 & AMAZE & 00:53:13.7 & +12:32:11.1 &  $3.212/7.54$ & 250 & $0.125\times0.25$ & [OIII] \\
    CDFS-9313 & AMAZE & 03:32:17.2 & -27:47:54.4 &  $3.654/7.21$ & 250 & $0.125\times0.25$ & [OIII] \\
    CDFS-9340$^{(4)}$ & AMAZE & 03:32:17.2 & -27:47:53.4 &  $3.658/7.20$ & 250 & $0.125\times0.25$ & [OIII] \\
    CDFS-11991 & AMAZE & 03:32:42.4 & -27:45:51.6 &  $3.661/7.25$ & 450 & $0.125\times0.25$ & [OIII] \\
    3C324-C3 & AMAZE & 15:49:47.1 & +21:27:05.0 &  $3.289/7.49$ & 150 & $0.125\times0.25$ & [OIII] \\
    DFS2237b-D29 & AMAZE & 22:39:32.7 & +11:55:51.7 & $3.370/7.43$ & 250 & $0.125\times0.25$ & [OIII] \\
    CDFS-5161 & AMAZE & 03:32:22.6 & -27:51:18.0 &  $3.66017.20$ & 300 & $0.125\times0.25$ & [OIII] \\
    DFS2237b-C21 & AMAZE & 22:39:29.0  & +11:50:58.0 & $3.403/7.40$ & 200 & $0.125\times0.25$ & [OIII] \\
    SSA22a-aug96M16 & AMAZE & 22:17:30.9  & +00:13:10.7 &  $3.292/7.48$ & 250 & $0.125\times0.25$ & [OIII] \\
    Q1422-D88 & AMAZE &14:24:37.9 & +23:00:22.3 &  $3.752/7.13$ & 250 & $0.125\times0.25$ & [OIII] \\
    CDFS-6664 & AMAZE & 03:32:33.3  & -27:50:7.4 &  $3.797/7.11$ & 500 & $0.125\times0.25$ & [OIII] \\
    SSA22a-C36 & AMAZE & 22:17:46.1 & +00:16:43.0 & $3.063/7.65$ & 100 & $0.125\times0.25$ & [OIII] \\
    CDFS-4414 & AMAZE & 03:32:23.2  & -27:51:57.9 &  $3.471/7.34$ & 350 & $0.125\times0.25$ & [OIII] \\
    CDFS-4417$^{(4)}$ & AMAZE & 03:32:23.3  & -27:51:56.8 &  $3.473/7.34$ & 350 & $0.125\times0.25$ & [OIII] \\
    CDFS-12631 & AMAZE & 03:32:18.1 & -27:45:19.0 &  $3.709/7.17$ & 250 & $0.125\times0.25$ & [OIII] \\
    CDFS-13497 & AMAZE & 03:32:36.3 & -27:44:34.6 &  $3.413/7.39$ & 150 & $0.125\times0.25$ & [OIII] \\
    CDFS-14411 & AMAZE & 03:32:20.9 & -27:43:46.3 &  $3.599/7.25$ & 200 & $0.125\times0.25$ & [OIII] \\
    CDFS-16272 & AMAZE & 03:32:17.1 & -27:42:17.8 &  $3.619/7.24$ & 350 & $0.125\times0.25$ & [OIII] \\
    CDFS-16767 & AMAZE & 03:32:35.9 & -27:41:49.9 &  $3.624/7.24$ & 300 & $0.125\times0.25$ & [OIII] \\
    Cosmic eye$^{(5)}$ & AMAZE & 21:35:12.7 & -01:01:42.9 &  $3.075/^{(5)}$ & 200 & $0.125\times0.25$ & [OIII] \\
    Abell1689-1$^{(5)}$ & AMAZE &  13:11:30.0 & -01:19:15.3 &  $3.770/^{(5)}$ & 300 & $0.125\times0.25$ & [OIII] \\
    Abell1689-2$^{(5)}$ & AMAZE &  13:11:25.5 & -01:20:51.9 &  $4.868/^{(5)}$ & 400 & $0.125\times0.25$ & [OII] \\
    Abell1689-4$^{(5)}$ & AMAZE & 13:11:26.5 & -01:19:56.8 &  $3.038/^{(5)}$ & 240 & $0.125\times0.25$ & [OIII] \\
    SSA22b-C5 & LSD & 22:17:47.1 & +00:04:25.7 &  $3.117/7.61$ & 240 & $0.05\times0.1$ (AO) & [OIII] \\
    SSA22a-C6 & LSD & 22:17:40.9 & +00:11:26.0 &  $3.097/7.63$ & 280 & $0.125\times0.25$ (AO) & [OIII] \\
    SSA22a-M4$^{(4)}$ & LSD & 22:17:40.9 & +00:11:27.9 &  $3.098/7.63$ & 280 & $0.125\times0.25$ (AO) & [OIII] \\
    SSA22a-C30 & LSD & 22:17:19.3 & +00:15:44.7 &  $3.104/7.62$ & 240 & $0.05\times0.1$ (AO) & [OIII] \\
    Q0302-C131 & LSD & 03:04:35.0 & -00:11:18.3 &  $3.240/7.51$ & 240 & $0.05\times0.1$ (AO) & [OIII] \\
    Q0302-C171 & LSD & 03:04:44.3 & -00:08:23.2 &  $3.34/7.44$ & 240 & $0.05\times0.1$ (AO) & [OIII] \\
    DSF2237b-D28 & LSD & 22:39:20.2 & +11:55:11.3 &  $2.938/7.75$ & 240 & $0.05\times0.1$ (AO) & [OIII] \\
    Q0302-M80 & LSD & 03:04:45.7 & -00:13:40.6 &  $3.416/7.39$ & 240 & $0.05\times0.1$ (AO) & [OIII] \\
    DSF2237b-MD19 & LSD & 22:39:21.1 & +11:48:27.7 & $2.616/7.99$ & 200 & $0.05\times0.1$ (AO) & H$\alpha$ \\
    \hline
  \end{tabular} 
\begin{list}{}{} 
\item[$(1)$] J2000.
\item[$(2)$] Redshifts are measured from the observed [OIII] line wavelength.
\item[$(3)$] [OIII] denotes the [OIII]$\lambda \lambda$ 5007,4959$\AA$ doublet; H$\alpha$ the H$\alpha\,\lambda$ 6563$\AA$ line.
\item[$(4)$] The object is in the same field of view of the object on the previous line
\item[$(5)$] Lensed objects not analyzed in this paper. The spatial scale ($pc/\arcsec$) depends on the lensing model.
\end{list}
\end{table*}

In this paper we focus on the dynamics of AMAZE and LSD  galaxies at $z\sim 3$.
In Sect.~\ref{s2} we present observations, data reduction and the method followed to extract kinematical maps from integral field spectra. In Sect.~\ref{s3} we present the results and a simple method to individuate the objects with a velocity map consistent with a rotating disk. In Sect.~\ref{s4} we describe the ``rotating disk'' model adopted in this paper, explain our ``fitting strategy'' to constrain model parameters and estimate their errors. In Sect.~\ref{s43} we discuss the results of model fitting on the selected subsample of objects consistent with rotating disks. In Sect.~\ref{s5} we discuss fit results and present the Tully-Fisher relation at $z\sim3$ (Sect.~\ref{s53}).

In the following, we adopt a $\Lambda CDM$ cosmology with $H_0=70 km s^{-1}Mpc^{-1}$, $\Omega _m=0.3$
and $\Omega _\Lambda=0.7$.

\section{Observations and data reduction}\label{s2}

Complete descriptions of the AMAZE and LSD programs, of their observations and data reduction are presented in \cite{Maiolino:2008} and \cite{Mannucci:2009}. Here we report a brief summary on observations and data reduction.

The near-IR spectroscopic observations were obtained by means of SINFONI, the integral field spectrometer at VLT \citep{Eisenhauer:2003}. For AMAZE galaxies, SINFONI was used in its
seeing-limited mode, with the $0.125\arcsec\times0.25\arcsec$ pixel scale and the H+K grism, yielding a spectral resolution $R\sim1500$ over the spectral range $1.45-2.41\mu m$. For the LSD galaxies, SINFONI was used with the Adaptive Optics module using a bright star close to the galaxy to guide the wavefront correction system. The pixel scale used is of $0.05\arcsec\times0.10\arcsec$ (for all object except {\it SSA22a-C6} and {\it SSA22a-M4} for which was used the  $0.125\arcsec\times0.25\arcsec$ pixel scale). The (K-band) seeing during the observations was generally about $0.6\arcsec-0.7\arcsec$. In the AO assisted observations the spatial resolution obtained is $\sim0.2\arcsec$.

Data were reduced by using the ESO-SINFONI pipeline (version 3.6.1). The pipeline subtracts the sky from the temporally contiguous frames, flat-fields the images, spectrally calibrates each individual slice and then reconstructs the cube. Individual cubes were aligned in the spatial direction using the offsets of the position of the $[OIII]$ or $H\alpha$ line emission peak.
Within the pipeline the pixels are resampled to a symmetric angular size of $0.125\arcsec \times 0.125\arcsec$ or $0.05\arcsec \times 0.05\arcsec$. The atmospheric absorption and instrumental response were taken into account and corrected by dividing with a suitable standard star.

In table \ref{tab1} we summarize all the relevant properties and observation setups of the object presented in this paper.

In the AMAZE sample there is a subsample of  4 lensed objects. We decided not to analyze these objects in this paper because of the uncertainties
introduced by the lensing model and the associated de-projection. They will be discussed in a forthcoming paper.

\subsection{Extraction of the gas kinematics}\label{s22}

We extract the kinematics of the gas by fitting the emission line spectrum for each spatial pixel of the cube corresponding to a given position on the sky.
In order to improve the signal-to-noise ratio (hereafter S/N) we first perform a smoothing of
the cube spatial planes, by using a gaussian filter with FWHM of 3 pixels. This is smaller than the spatial resolution of the observations which usually corresponds to a Point Spread Function with a FWHM of at least 4 pixels.

In all cases, but one, we fit the profile and shift of [OIII]$\lambda\lambda 5007, 4959$\AA\
doublet. The two [OIII] lines are parametrized using single gaussian functions with the
following, physically motivated constraints:  the two lines are forced to have the same average velocity $v$ and velocity dispersion $\sigma$, while their flux ratio is fixed, $F_{4959}/F_{5007}= 0.33$, since both the lines are emitted from the same upper energy level of the $O^{2+}$ ion.
For one object at $z\sim 2.6$ we fit the $H\alpha$ line with a single gaussian function.
Since we perform the fit of the spectrum within a spectral window of only $\sim 0.2\mu m$
around the line, the continuum level is simply set to the mean continuum flux in this band. 

To automatically exclude noise fluctuations or bad pixels and to take into account the effect of instrumental broadening we constrain the velocity dispersion of the lines to be larger than the instrumental resolution estimated from the sky emission lines. We define the signal-to-noise ratio (S/N) of each line as the peak of the line model divided by the r.m.s. of the spectrum estimated in regions with no emission lines. Finally, we reject all fits where the S/N is lower than $2$.

Uncertainties on measured fluxes, velocities and velocity dispersions are obtained from the formal errors on best fit parameters. Such errors are computed after two iterations of the line fitting procedure: in the second fit we adopt pixel by pixel errors equal to the r.m.s. of the residuals in the first fit.

\section{Results}\label{s3}

  \begin{figure*}[!ht]
  \centering
  \includegraphics[width=0.84\linewidth]{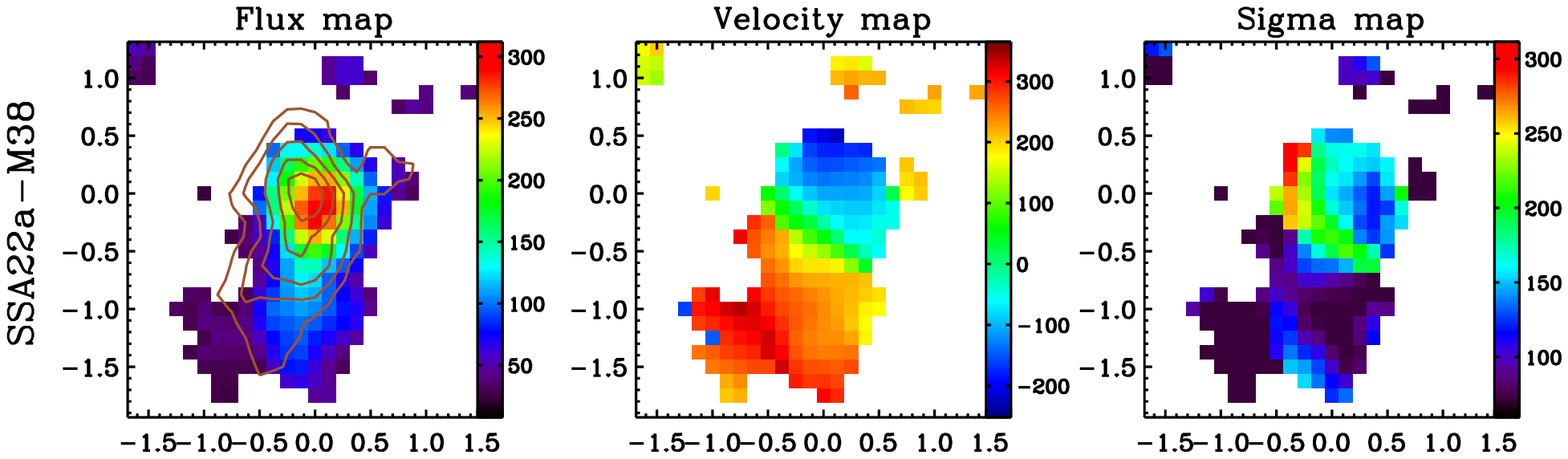}\\
  \includegraphics[width=0.84\linewidth]{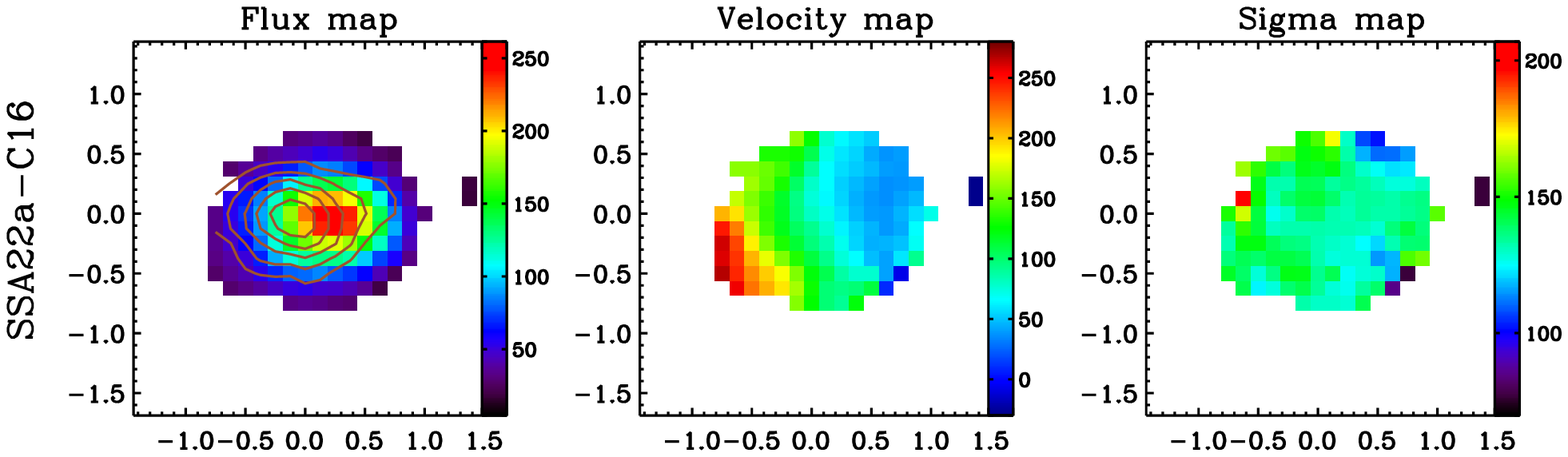}\\
  \includegraphics[width=0.84\linewidth]{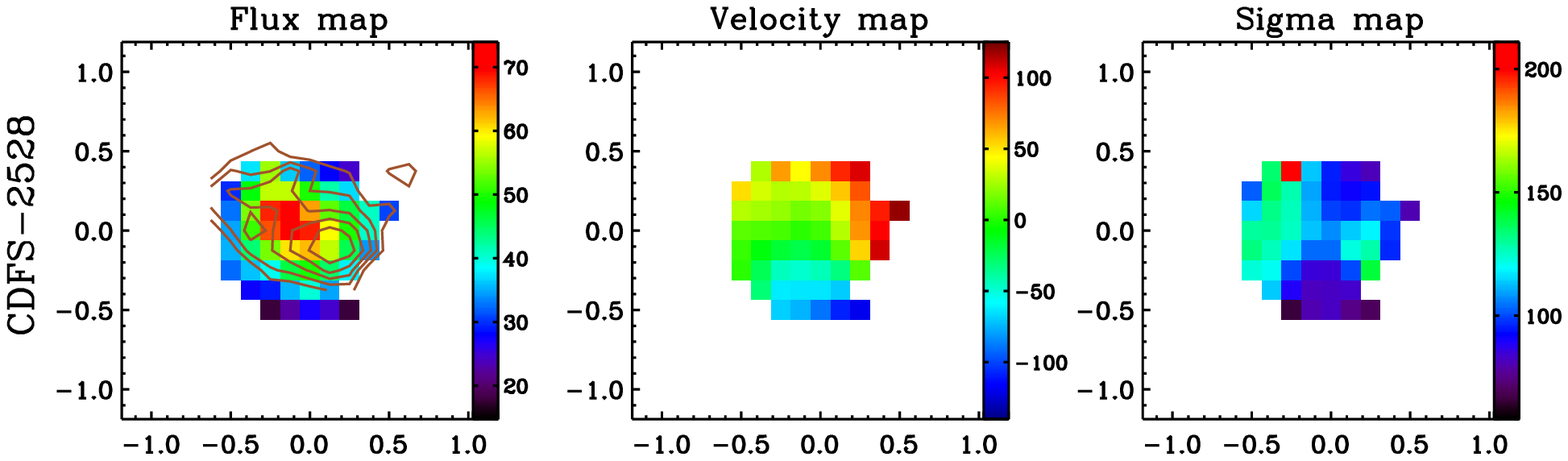}\\
  \includegraphics[width=0.84\linewidth]{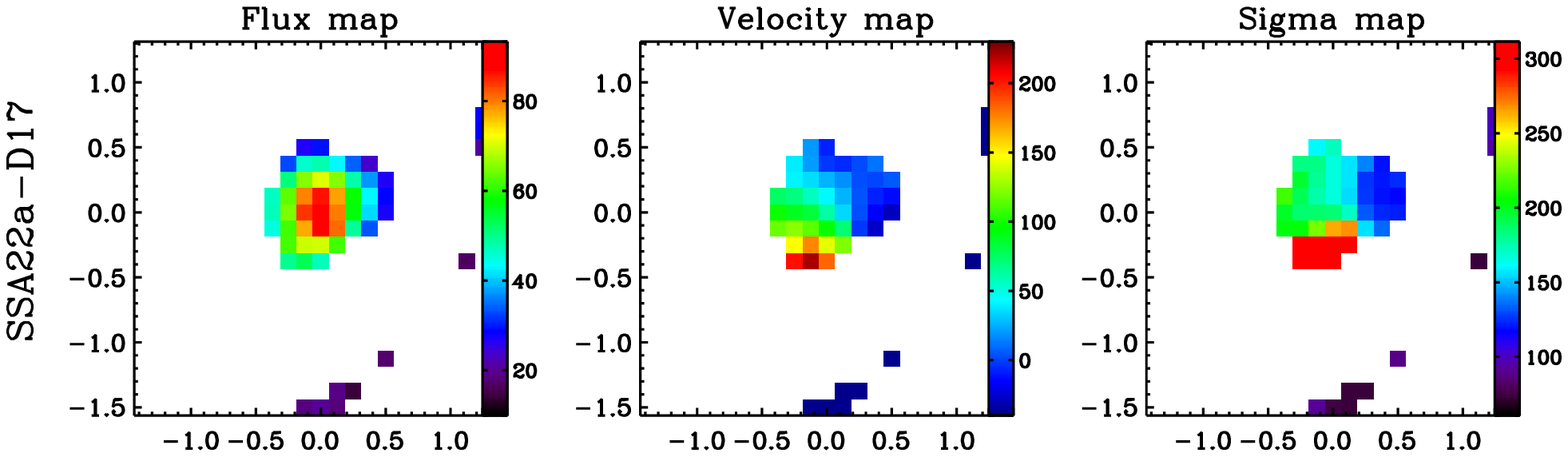}\\
 \caption{Kinematical maps for objects {\it SSA22A-M38}, {\it SSA22A-C16},  {\it CDFS-2528} and {\it SSA22A-D17}  (from the top respectively). Respectively, from the left: flux, velocity and sigma map. the $X-Y$ coordinates are in arcseconds referred to an arbitrary object centre position. The North direction is the positive Y axis. The vertical color bars are in arbitrary units for the flux map and in $km\ s^{-1}$ for the velocity and sigma maps. Overplotted on the flux map the continuum flux distribution (brown isophotes) for the object whenever we detect a continuum flux component with sufficient S/N.}
  \label{figa1}
  \end{figure*}
  
  \begin{figure*}[!ht]
  \centering
  \includegraphics[width=0.84\linewidth]{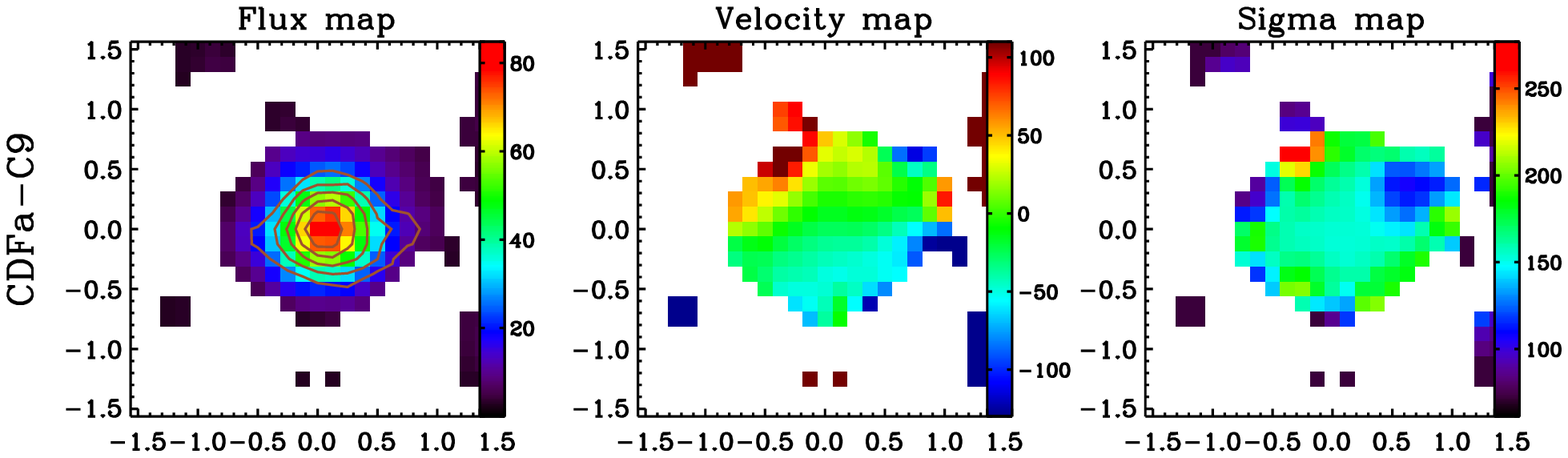}\\
  \includegraphics[width=0.84\linewidth]{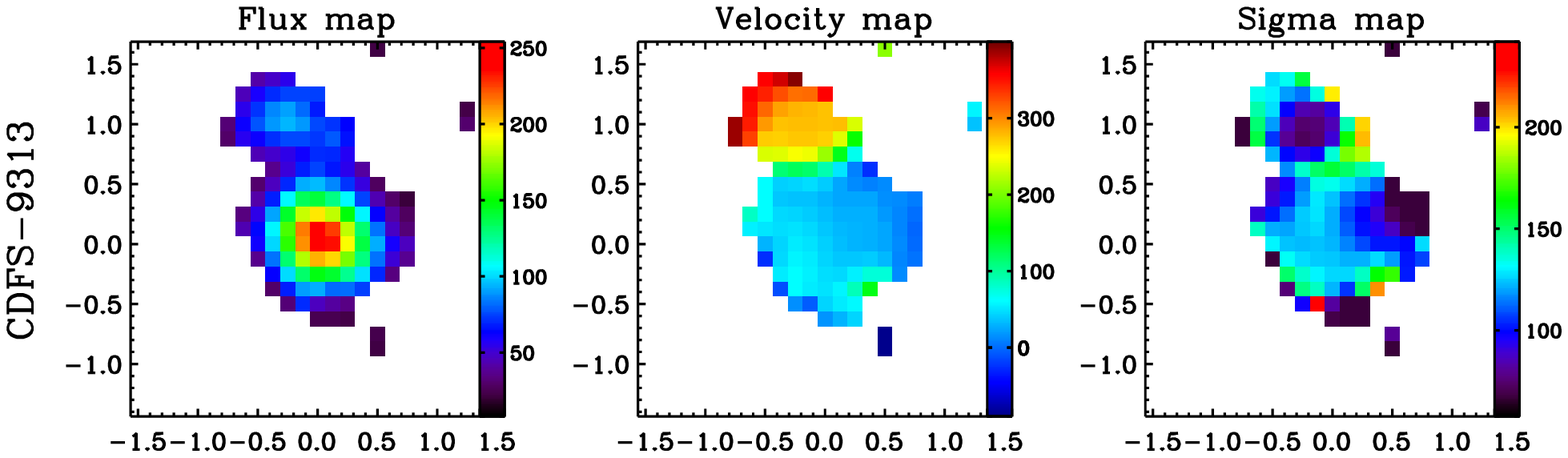}\\
  \includegraphics[width=0.84\linewidth]{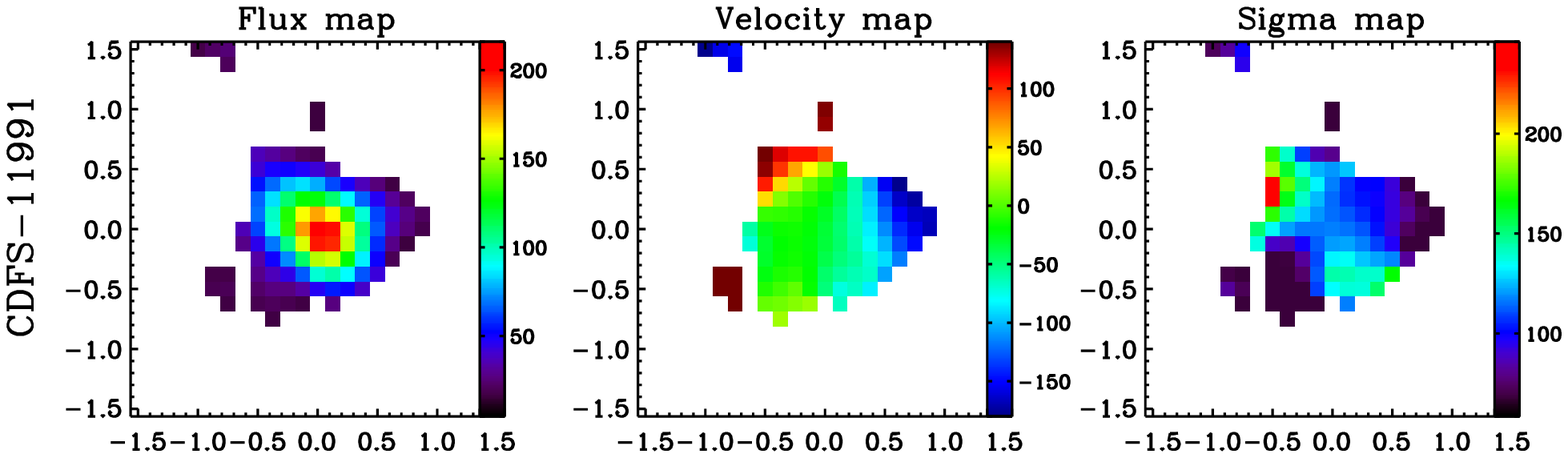}\\
  \includegraphics[width=0.84\linewidth]{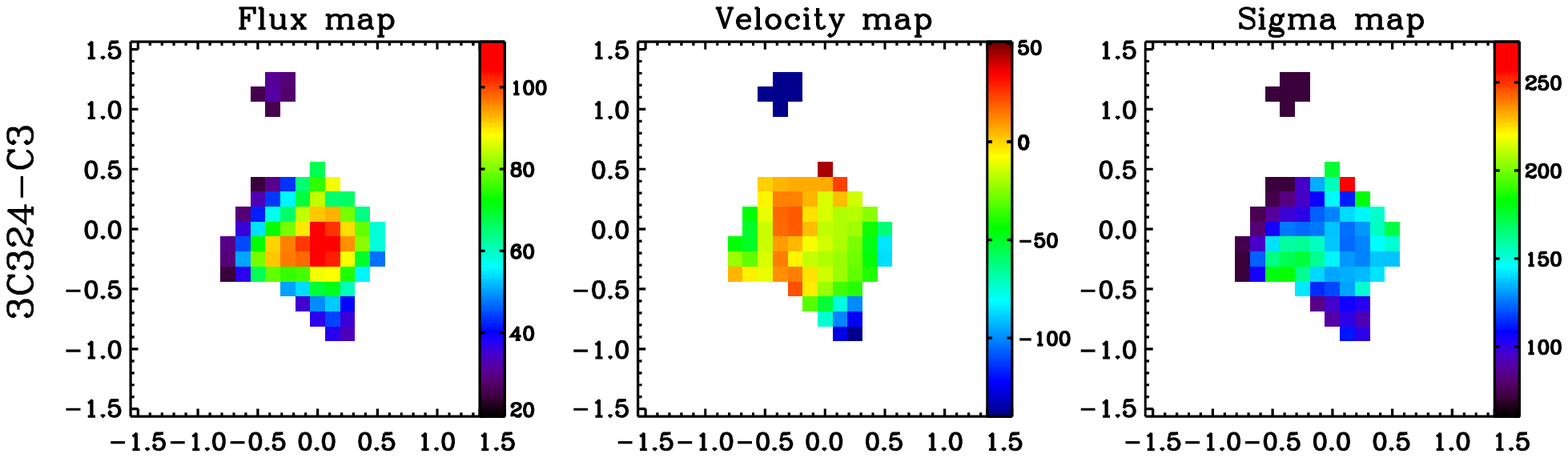}\\
  \includegraphics[width=0.84\linewidth]{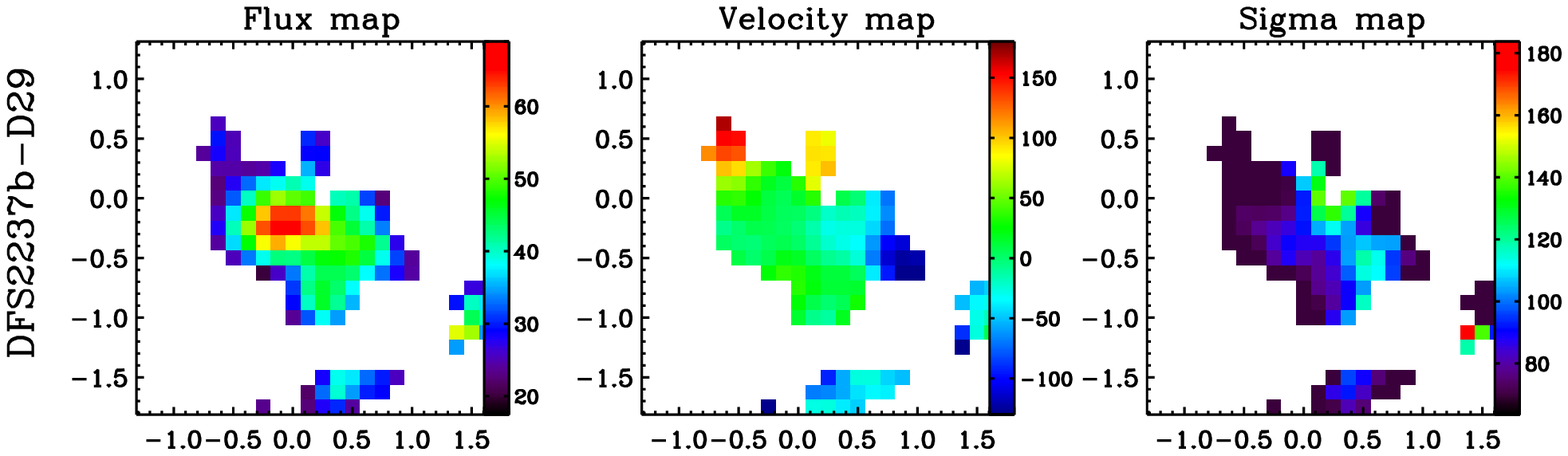}
 \caption{Kinematic maps for the objects (respectively from the top) {\it CDFa-C9},  {\it CDFS-9313} (in the same field of view the fainter source at the North-West is {\it CDFS-9340}), {\it CDFS-11991}, {\it 3C324-C3} and {\it DFS2237b-D29}. Panels, axis and color bar indications same as in Fig.~\ref{figa1}.}
   \label{figa2}
  \end{figure*}
    
  \begin{figure*}[!ht]
  \centering
  \includegraphics[width=0.84\linewidth]{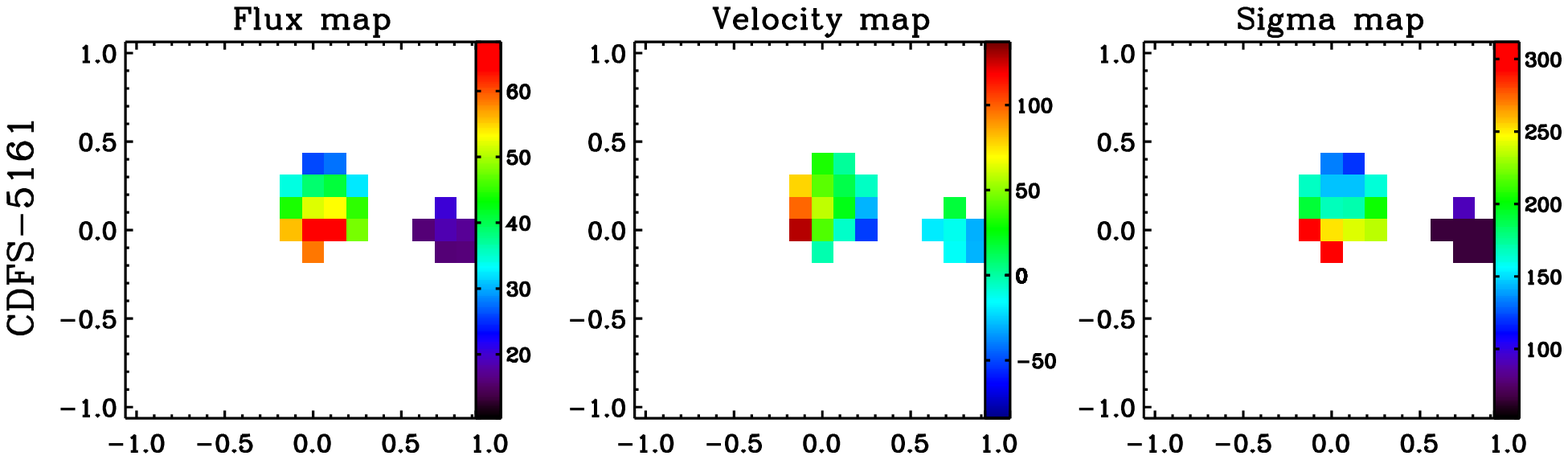}\\
  \includegraphics[width=0.84\linewidth]{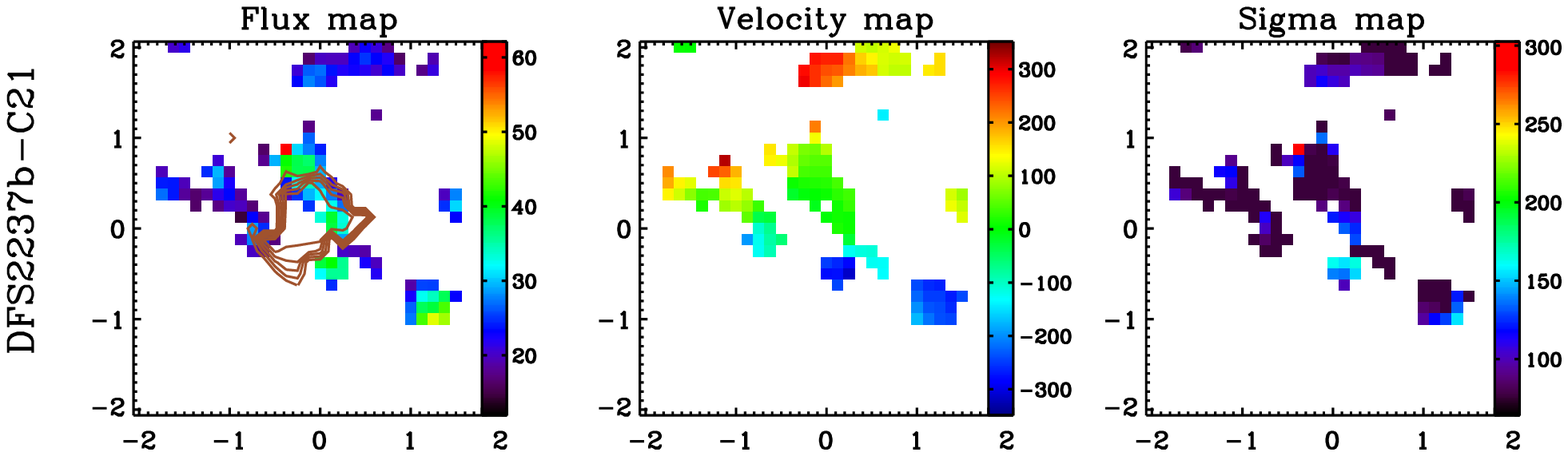}\\
  \includegraphics[width=0.84\linewidth]{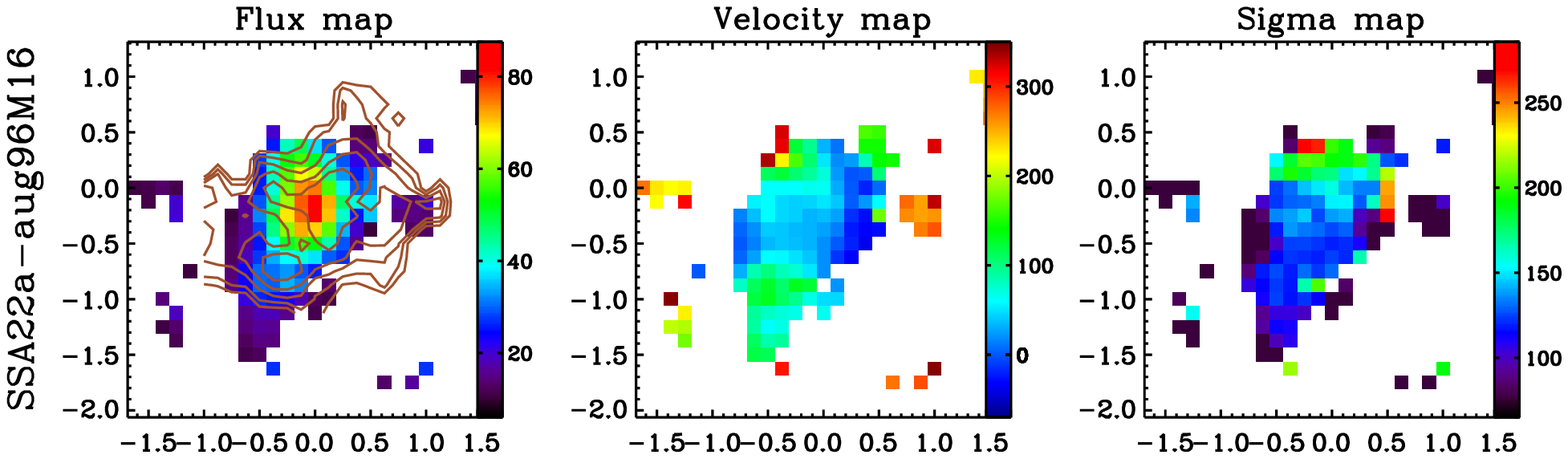}\\
  \includegraphics[width=0.84\linewidth]{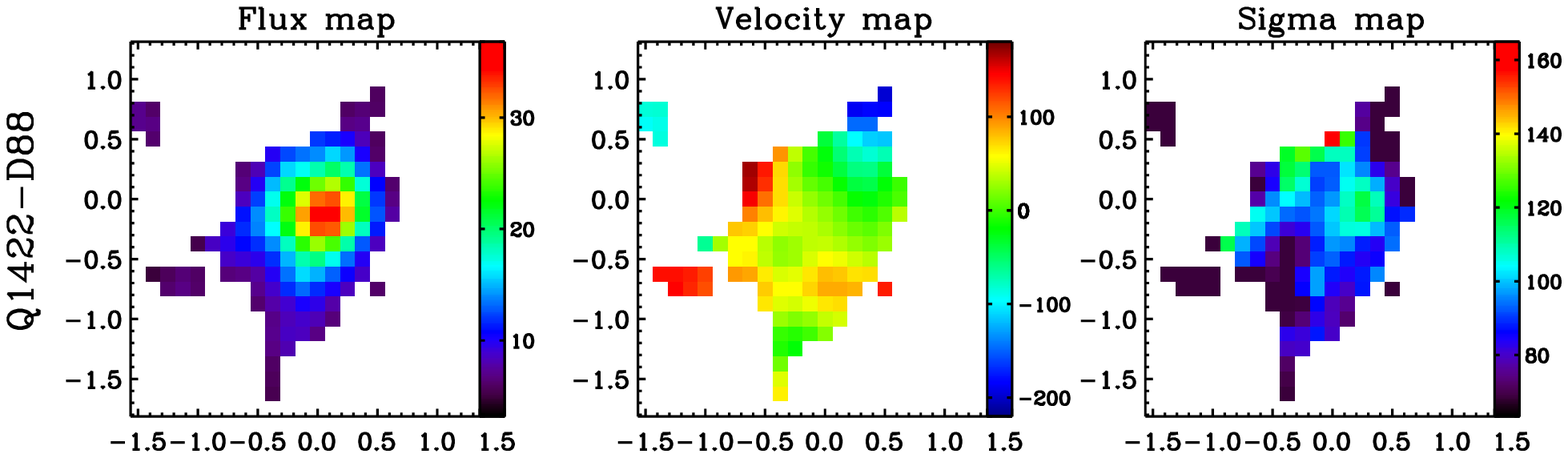}\\
  \includegraphics[width=0.84\linewidth]{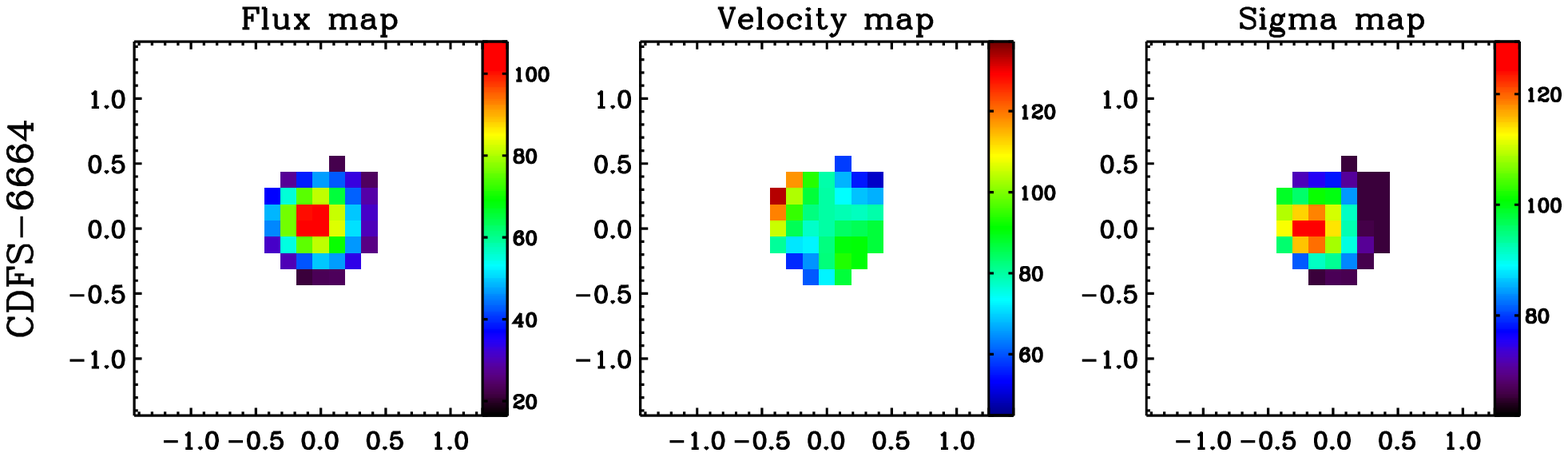} 
    \caption{Kinematic maps for the objects (respectively from the top) {\it CDFS-5161},  {\it DFS2237b-C21}, {\it SSA22a-aug96M16}, {\it Q1422-D88} and {\it CDFS-6664}. Panels, axis and color bar indications same as in Fig. \ref{figa1}.}
  \label{figa3}
  \end{figure*}
  
  \begin{figure*}[!ht]
  \centering
  \includegraphics[width=0.84\linewidth]{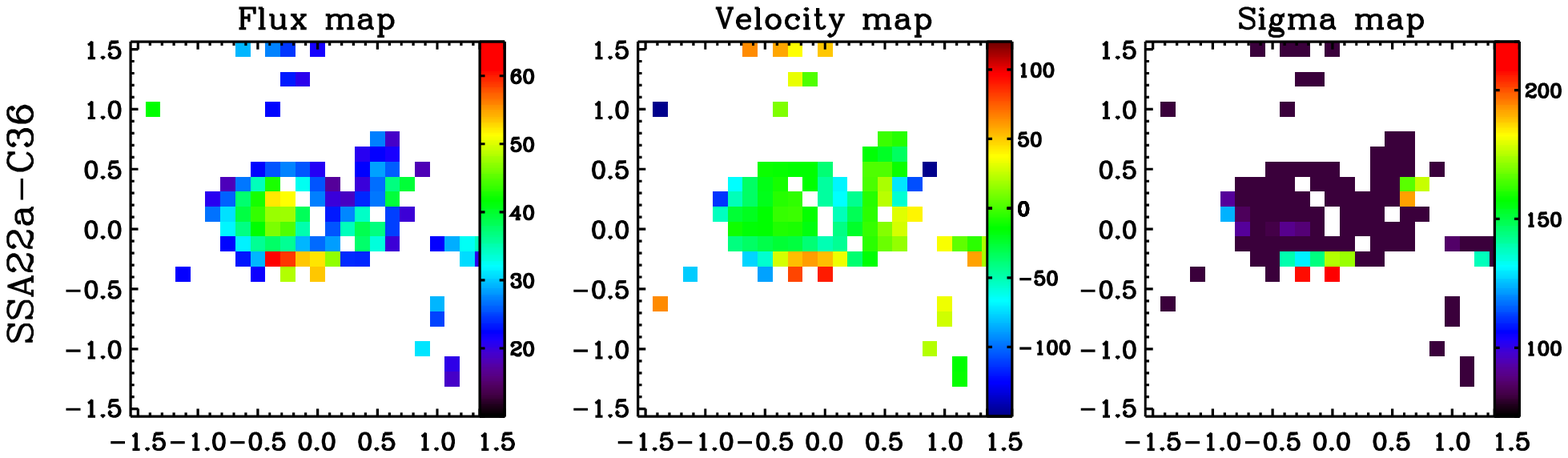}\\
  \includegraphics[width=0.84\linewidth]{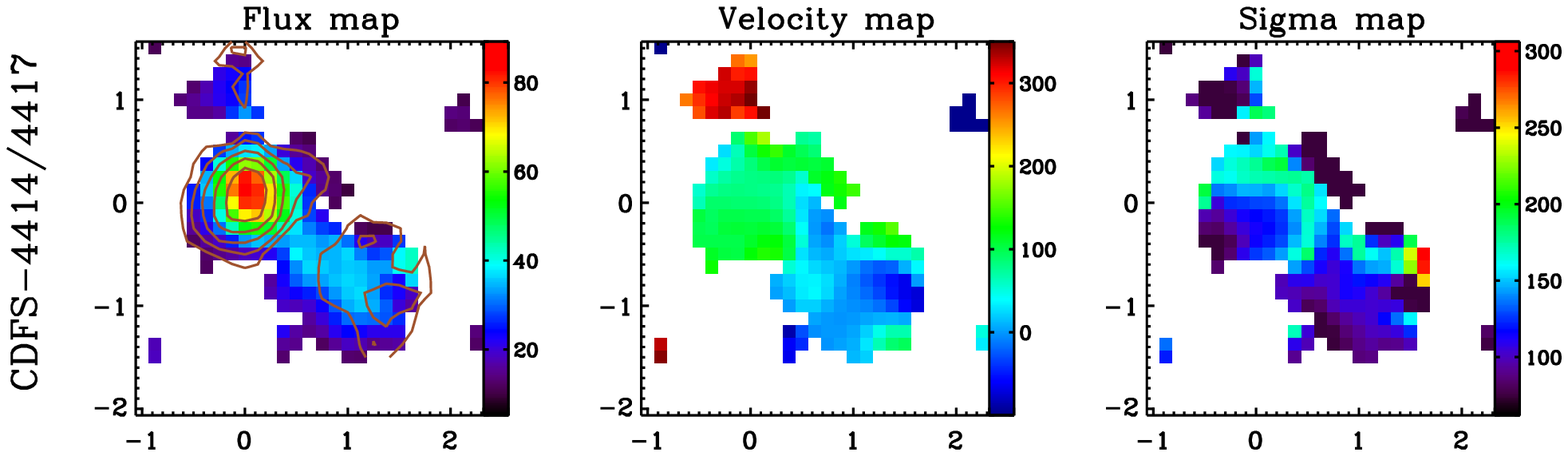}\\
  \includegraphics[width=0.84\linewidth]{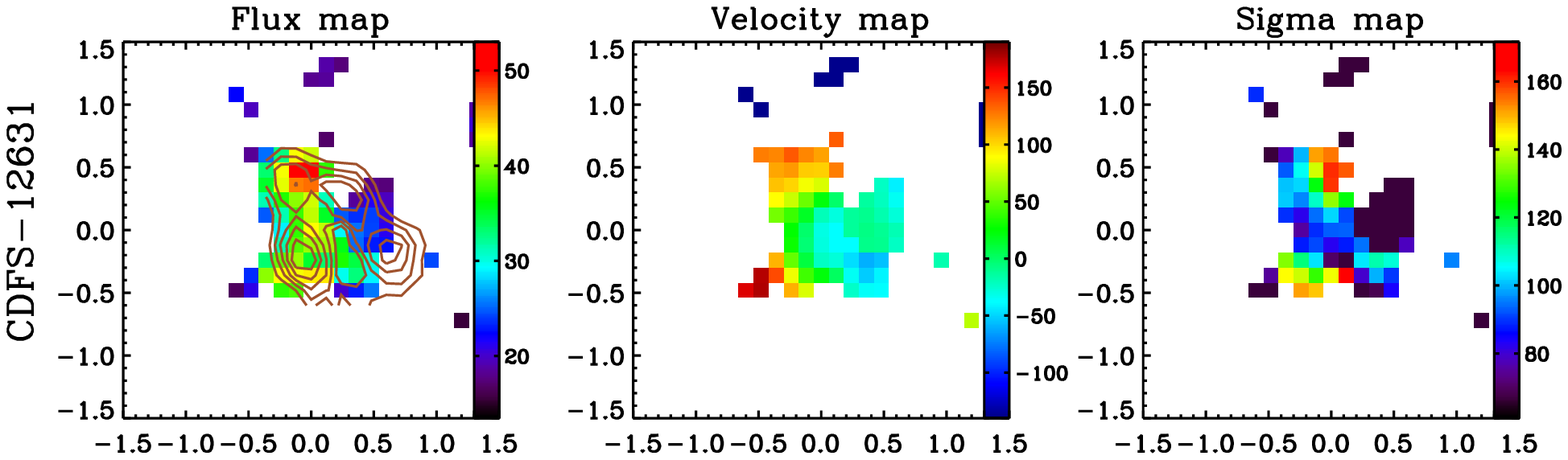}\\
  \includegraphics[width=0.84\linewidth]{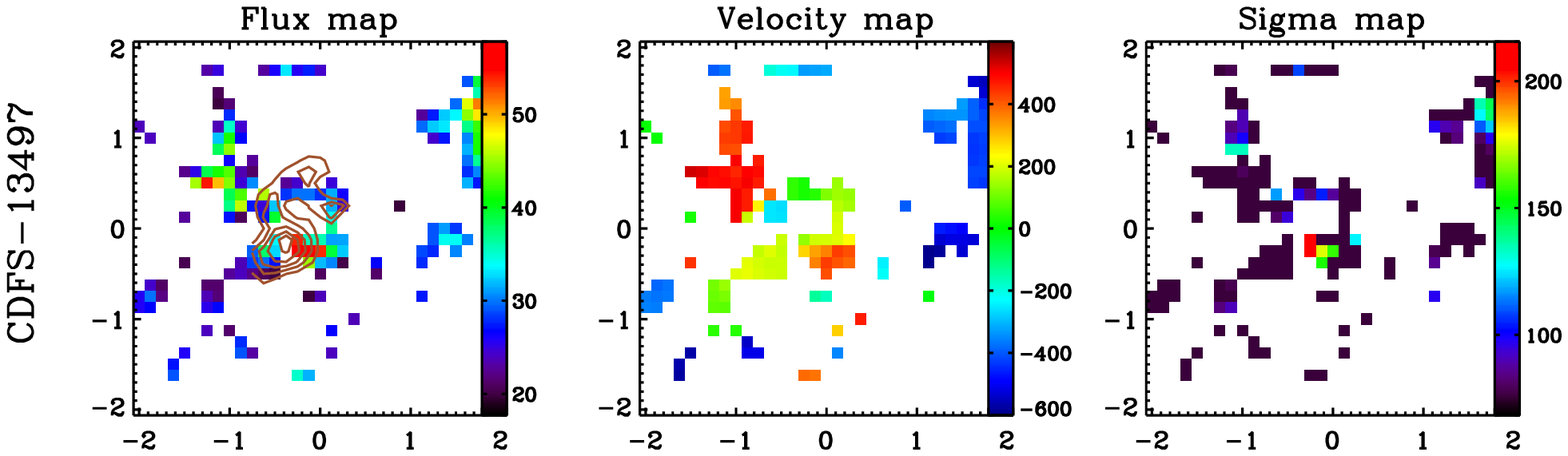}\\
  \includegraphics[width=0.84\linewidth]{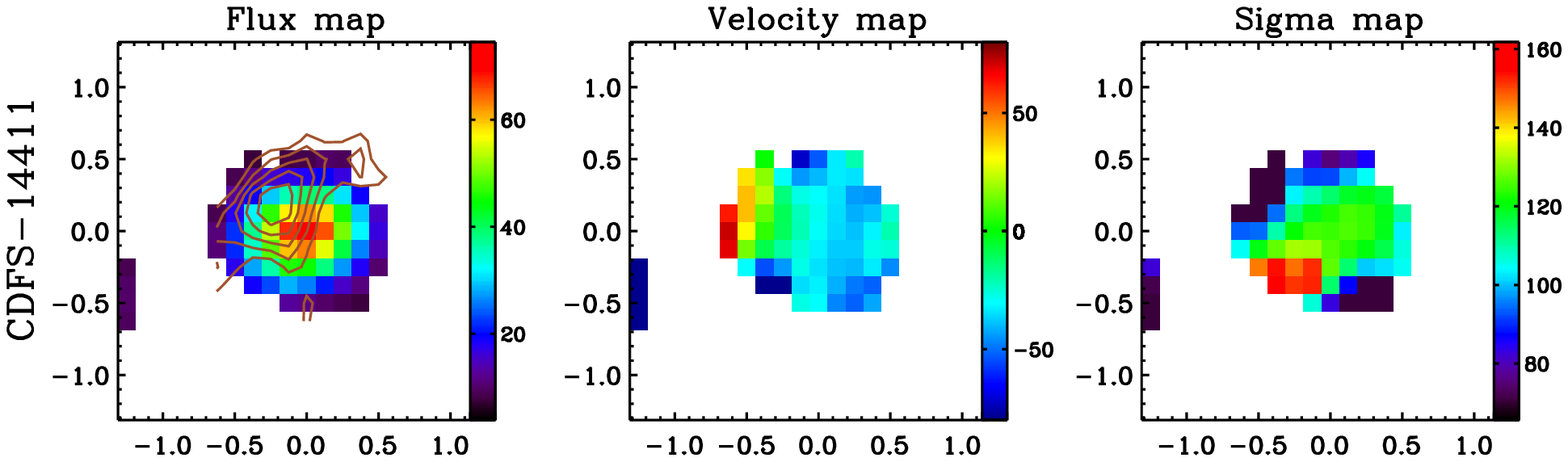} 
 \caption{Kinematic maps for the objects (respectively from the top) {\it SSA22A-C36}, {\it CDFS-4414/4417}({\it CDFS-4417} is the brighter object at the north), {\it CDFS-12631},{\it CDFS-13497} and {\it CDFS-14411}. Panels, axis and color bar indications same as in Fig. \ref{figa1}.}
  \label{figa4}
  \end{figure*}
  
  \begin{figure*}[!ht]
  \centering
  \includegraphics[width=0.84\linewidth]{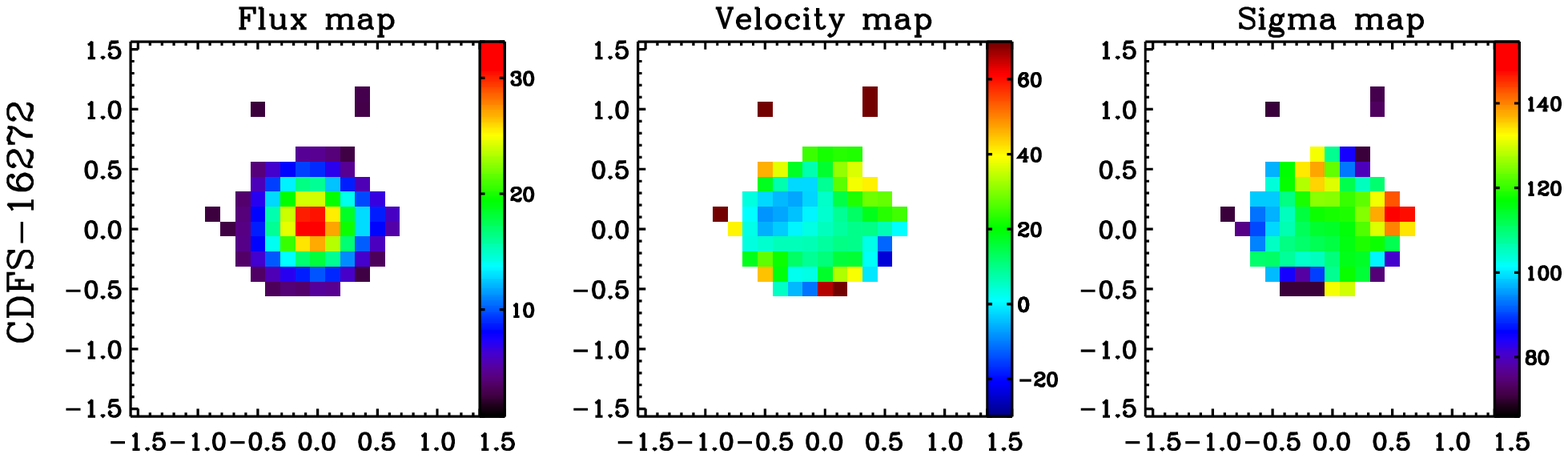}\\
  \includegraphics[width=0.84\linewidth]{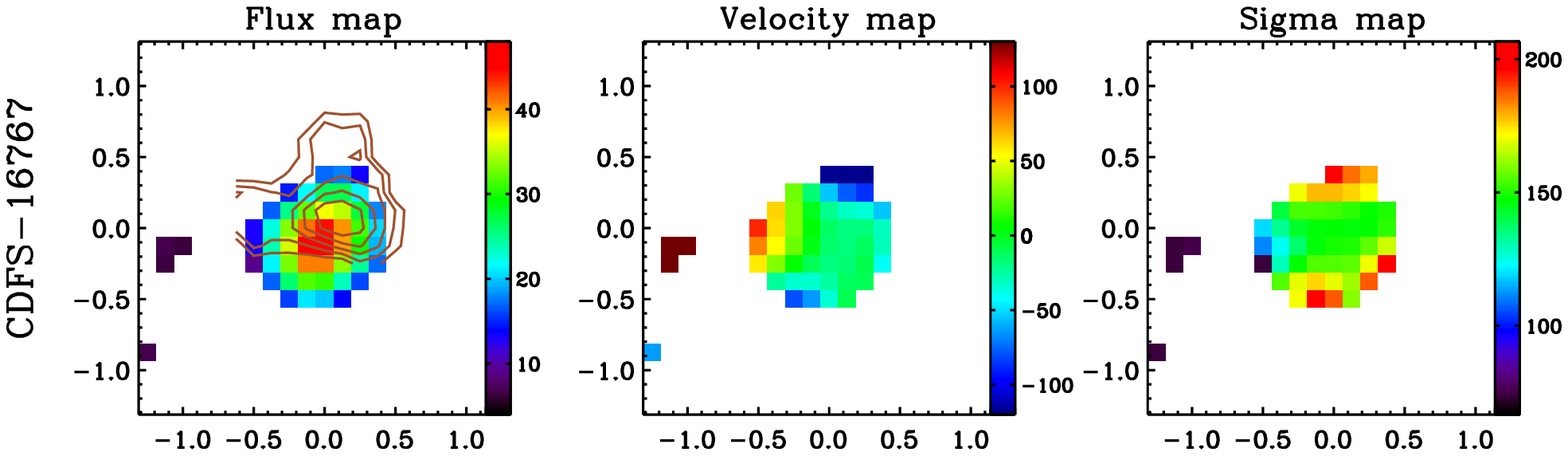}\\
  \includegraphics[width=0.84\linewidth]{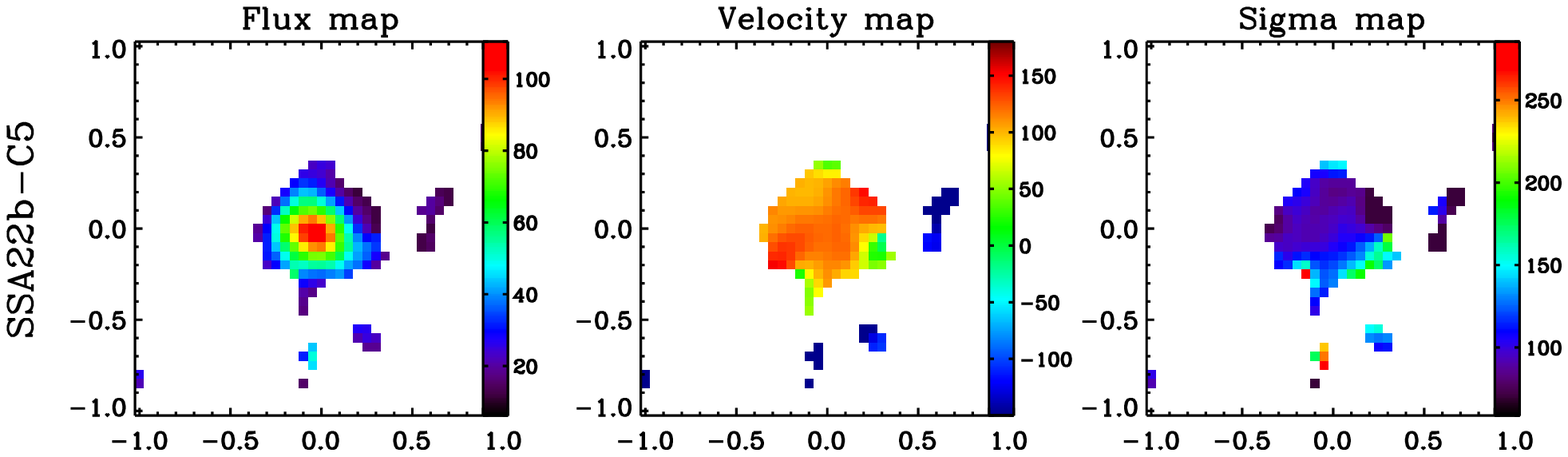}\\
  \includegraphics[width=0.84\linewidth]{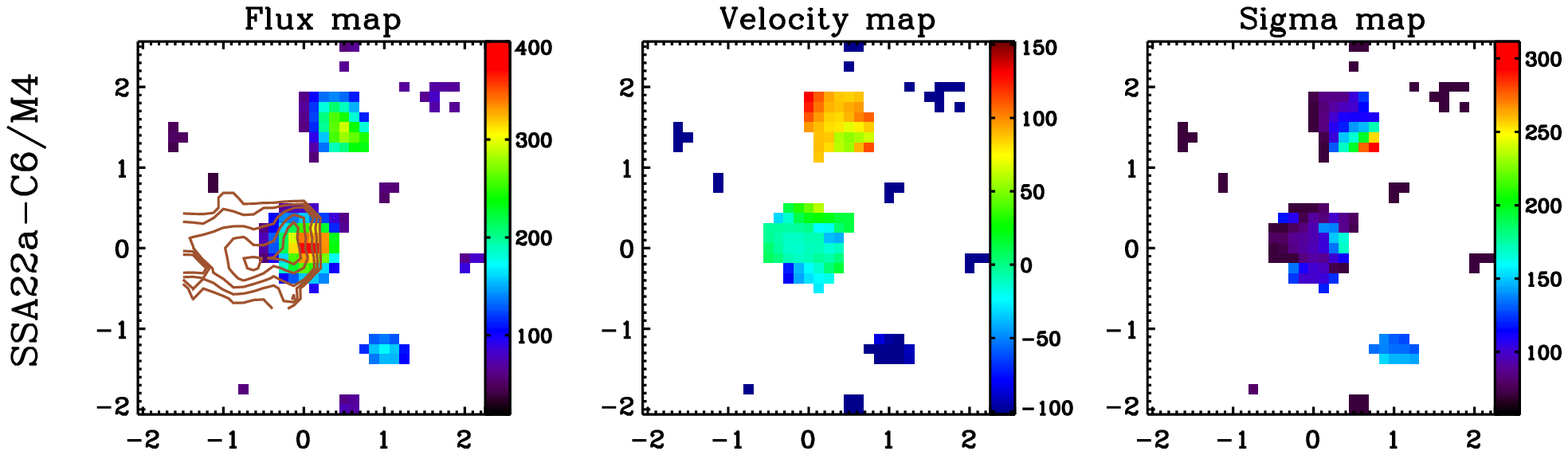}\\
  \includegraphics[width=0.84\linewidth]{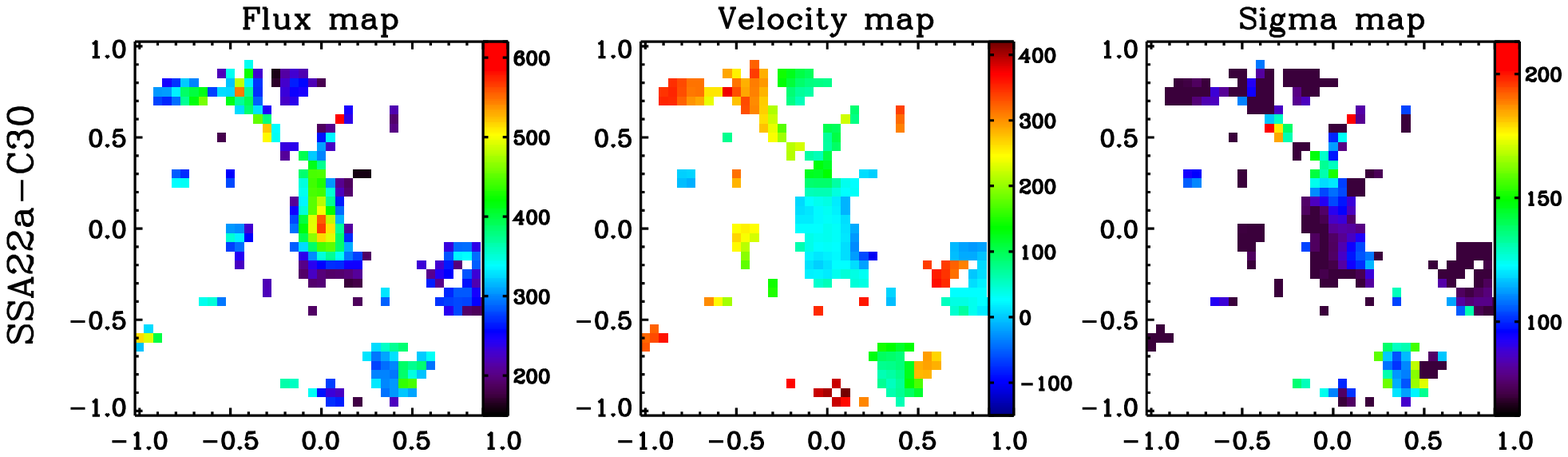}
  \caption{Kinematic maps for the objects (respectively from the top) {\it CDFS-16272},  {\it CDFS-16767},  {\it SSA22b-C5}, {\it SSA22a-C6/M4} ({\it SSA22a-M4} is the fainter northern object) and {\it SSA22a-C30}. Panels, axis and color bar indications same as in Fig. \ref{figa1}.}
  \label{figa5}
  \end{figure*}
    
  \begin{figure*}[!ht]
  \centering
  \includegraphics[width=0.84\linewidth]{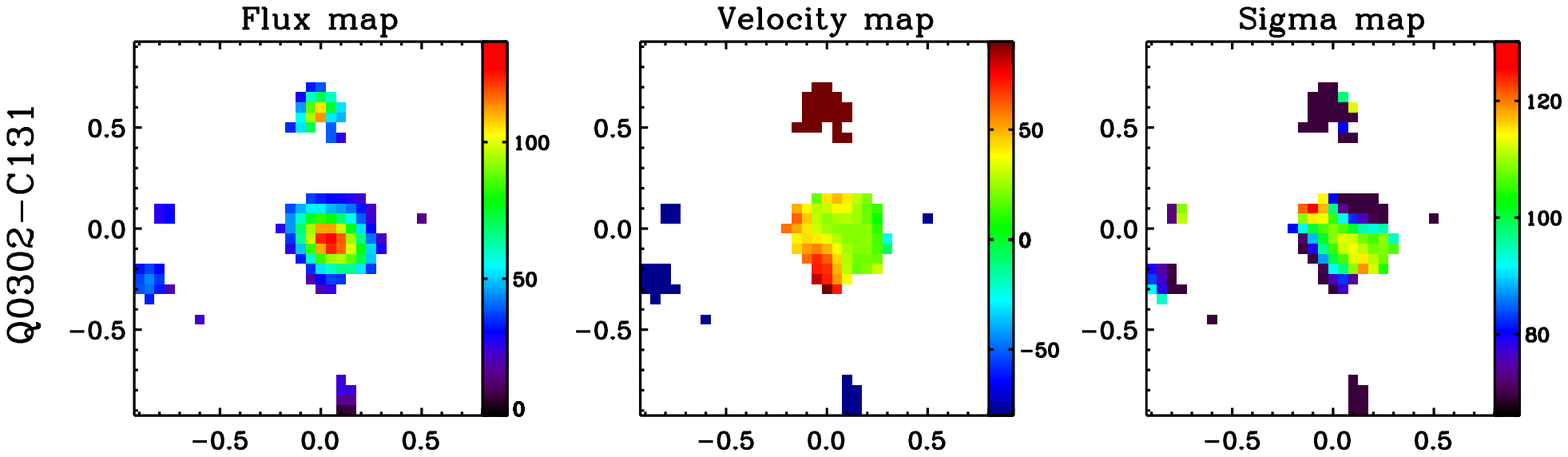}\\
  \includegraphics[width=0.84\linewidth]{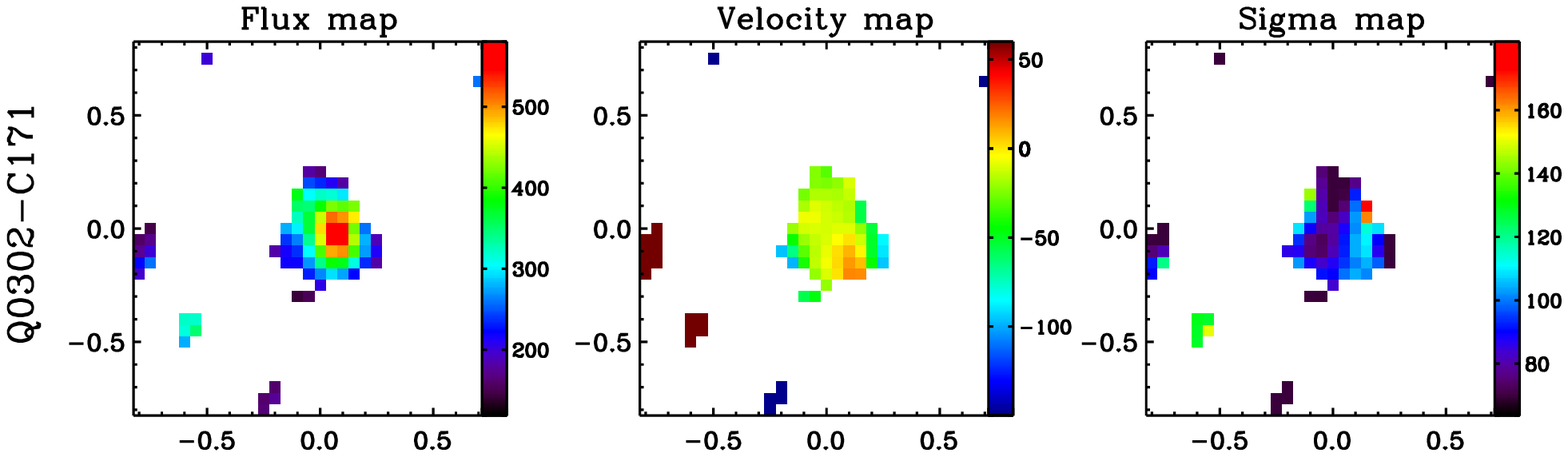}\\
  \includegraphics[width=0.84\linewidth]{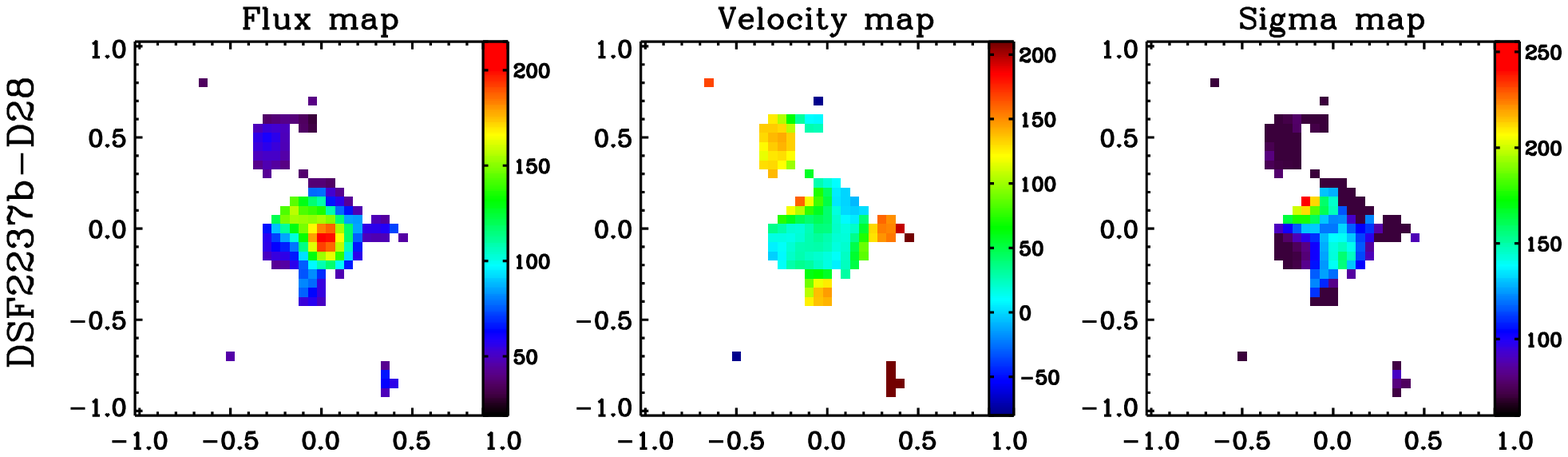}\\
  \includegraphics[width=0.84\linewidth]{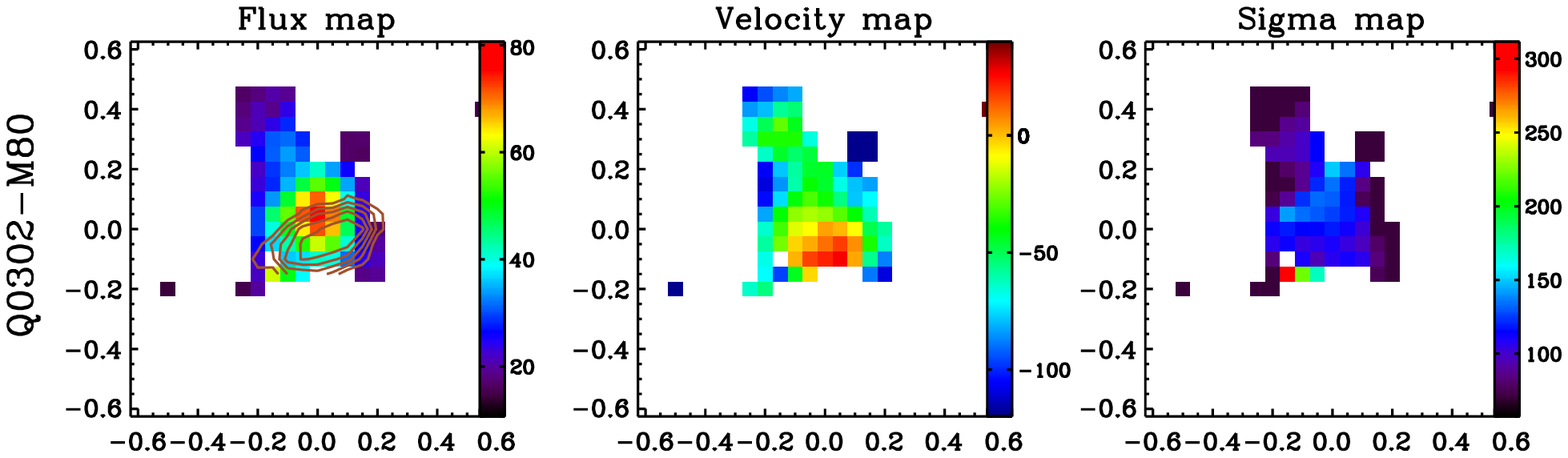}\\
  \includegraphics[width=0.84\linewidth]{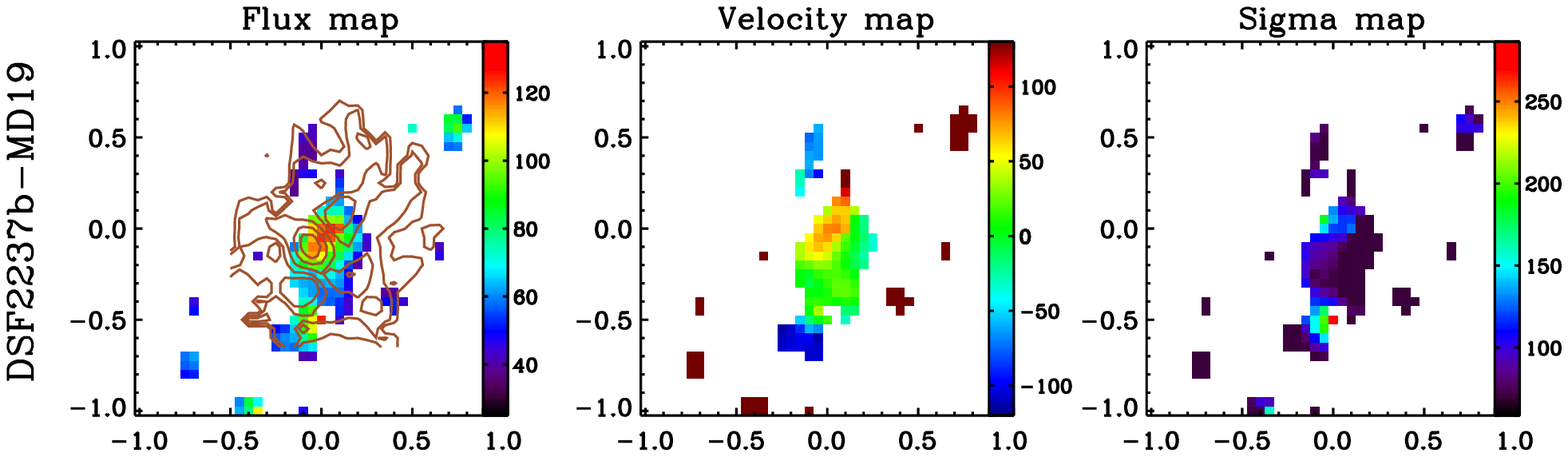}
    \caption{Kinematic maps for the objects (respectively from the top)  {\it Q0302-C131},  {\it Q0302-C171}, {\it DSF2237b-D28},  {\it Q0302-M80} and {\it DSF2237b-MD19}. Panels, axis and color bar indications same as in Fig. \ref{figa1}.}
  \label{figa6}
  \end{figure*}

In Fig.~\ref{figa1} to \ref{figa6} we present the kinematical maps for all the objects
described in tab.~\ref{tab1}. This paper is dedicated only to the study of galaxy dynamical
properties while other studies (e.g.~integrated properties,
metallicity, etc.) have been or will be presented in other dedicated papers (\citealt{Maiolino:2008}, \citealt{Mannucci:2009}, \citealt{Mannucci:2010}, \citealt{Cresci:2010}~and Troncoso et al.~2010, in prep.).

In the majority of the AMAZE sample (17 out of 23 objects, i.e. $\sim72\%$,
excluding the four lensed objects) the morphology of line flux maps is simple, 
characterized by single peak distributions with typical dimensions of $\sim1-1.5\arcsec$ and
relatively low ellipticities of the isophotes. These morphologies are likely the consequence of
beam smearing since the typical seeing in this data is $\sim0.6-0.7\arcsec$ FWHM (see below for
a more accurate estimation). However, in some cases we find a secondary clump or asymmetric
extended emission (e.g. in SSA22-M38). 

In the galaxies from the LSD sample we find more complex morphologies which are likely the consequence of the AO-assisted observations providing higher spatial resolution ($\sim0.2\arcsec$ FWHM) which allows us to spatially resolve more complex structures, but lower sensitivity to extended sources given the higher surface brightness detection threshold.

There are three cases of close pairs of objects with similar brightnesses. {\it CDFS-9313} and {\it CDFS-9340} have a projected plane of the sky separation of $\sim1.0\arcsec$ ($\sim7kpc$ at the average redshift of the sources); {\it CDFS-9340} is redshifted by $\sim280kms^{-1}$ with respect to {\it CDFS-9313}, the brightest of the pair. {\it CDFS-4414} and {\it CDFS-4417} have a projected separation of $\sim1.2\arcsec$ ($\sim9kpc$); {\it CDFS-4414} is blueshifted by $\sim120kms^{-1}$ with respect to the brighter companion. {\it SSA22a-C6} and {\it SSA22a-M4}, from the LSD sample, have a projected separation of $\sim1.6\arcsec$ ($\sim12kpc$) and {\it SSA22a-M4} is redshifted by $\sim90kms^{-1}$ with respect to the brighter companion.

Inspection of the velocity maps in Fig.~\ref{figa1} to \ref{figa6}  reveals a number of objects with a clear and regular velocity gradient (12 out of 22 objects in the AMAZE sample and only one out of 9 in the LSD sample).
The presence of such velocity gradients is an indication of possible rotating-disk kinematics. 
The identification of rotating objects and their distinction from complex kinematics cases or mergers is important for the comparison with theoretical models.
 Some authours (\citealt{Forster-Schreiber:2009}, \citealt{Cresci:2009}) used the technique developed
 by \cite{Shapiro:2008} based on kinemetry \citep{Krajnovic:2006} to quantify asymmetries in
 both the velocity and sigma maps in order to empirically differentiate between rotating and
 non rotating systems. This technique is not useful when applied to the data presented in this
 paper due to the low S/N that does not allow us to recover with sufficient accuracy the needed kinemetry parameters.
We assess the presence of such velocity gradients by using a different approach. 
We fit the velocity map with a plane (i.e. $v=ax+by+v_0$) and classify the object as
``rotating'' or ``not rotating'' if the velocity map is well fitted (or not) by a plane. In
detail we use the $\chi^2$ statistic to accept or reject the hypothesis that the velocity map
is well fitted by a plane: we accept the hypothesis if there is a probability lower than the
$4.6\%$ (adopting a $\chi^2$ distribution with the correct number of degree of freedom) to obtain a $\chi^2$ value greater than the value obtained.

For objects with velocity maps well fitted by a plane we calculate its inclination relative
to the $X-Y$ plane (i.e the magnitude of the velocity gradient). We confirm its ``rotating''
classification only if this inclination is inconsistent with zero at least a significance level of $4 \sigma$. On the other hand a plane inclination consistent with zero might correspond either to genuinely non rotating object or to a rotating one seen face-on. In this case we cannot identify it as a rotating disk. 
In conclusion we classify the observed galaxies with the following criteria.
\begin{itemize}
\item If the velocity map shows a non zero gradient from plane fitting, we classify the object as ``rotating''.
\item If the velocity map is well fitted by a plane but its inclination is consistent with zero within $4 \sigma$, we label the object as ``not classifiable''.
\item If the velocity map cannot be fitted with a plane, we classify the object as ``not rotating''.
\end{itemize}

We choose to fit a ``plane'' to the velocity gradient because, due to the limited S/N, we
are sampling preferentially the linear part of the rotation curve of our targets. By using
a plane we can adopt a simple model with few free parameters and therefore
we can rely on more simple statistics.

In principle we could have used also velocity dispersion maps to classify galaxies as ``rotating'' or ``non rotating'': in fact, a circularly symmetric dispersion map is the expected signature of spatially unresolved rotation at the center of a rotating disk. However, we decided not to use them principally because we would have missed galaxies with regular rotation patterns in velocity, but whose velocity dispersion maps are distorted by turbulent and/or non-gravitational motions, as verified ``a posteriori'' with our modeling.

\begin{figure}[!ht]
  \centering
  \includegraphics[width=0.99\linewidth]{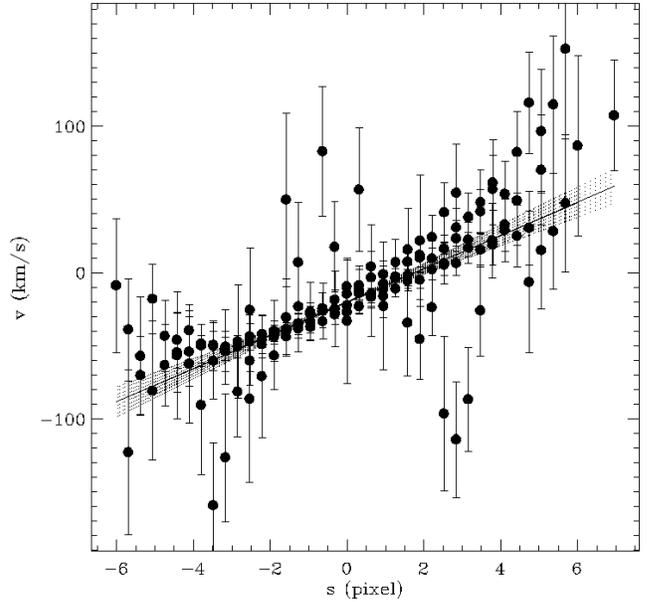}
  \caption{Example of the method (applied to object {\it CDFA-C9}) for assessing the presence of a smooth gradient in the velocity map. The solid line shows the fitted plane seen from the edge, while the filled points show the observed velocity map on the projection. The dashed lines represent the $1$,$2$,$3$, and $4\sigma$ on the plane inclination.}
  \label{fig1x}
  \end{figure}

In Fig.~\ref{fig1x} we show an example of our method. We plot the projection of the velocity
map along the direction ($s$) of maximum inclination of the fitted plane (i.e. the fitted plane
seen from its edge). We also show the $1$,$2$,$3$, and $4\sigma$ slopes of the plane inclination.

The classification for all the objects is reported in table \ref{tab2}.

Note that the object {\it SSA22a-M38} is classified as ``rotating" even if its velocity
map is not fitted by a plane (see the $\chi^2$ value in tab.~\ref{tab2}). In this case the
smooth velocity gradient is very clear and the S/N is high; the fit with a simple plane fails because of the flattening of the velocity field at the ``red'' and ``blue'' edges.

The next step is to model the gas kinematics of ``rotating'' galaxies to relate observed kinematics with the mass distribution.

\begin{table}[!ht]
  \caption[!ht]{Objects observed.}
  \label{tab2}
  \centering
  \begin{tabular}{l c c}
    \hline
    \hline
    \noalign{\smallskip}
    AMAZE Objects  &  $\chi^2/d.o.f.$ & Classification\\
    \hline
    \noalign{\smallskip}
    SSA22a-M38 & $3057/176$ & Rotating\\ 
    SSA22a-C16 &  $182/136$ & Rotating\\ 
    CDFS-2528 &  $61.6/52$ & Rotating\\
    SSA22a-D17 & $20.6/42$  & Rotating\\
    CDFA-C9 & $157/142$  & Rotating\\
    CDFS-9313  & $63.8/91$  & Rotating\\
    CDFS-9340  & $42.5/29$  & Rotating\\
    CDFS-11991 &$226/99$   &  {\tiny Not rot.}\\
    3C324-C3 & $45.3/76$  & Rotating\\
    DFS2237b-D29 & $229/107$  &  {\tiny Not rot.}\\
    CDFS-5161 & $2.00/12$   & Not clas.\\ 
    DFS2237b-C21 & $725/141$   &  {\tiny Not rot.}\\
    SSA22a-aug96M16 & $680/119$   &  {\tiny Not rot.}\\
    Q1422-D88 &  $478/127$   &  {\tiny Not rot.}\\
    CDFS-6664 &  $15.9/40$  &  Not clas.\\
    SSA22a-C36 & $102/86$   &  Not clas.\\
    CDFS-4414 & $259/91$  &  {\tiny Not rot.}\\
    CDFS-4417 & $117/94$  &  Not clas.\\
    CDFS-12631 &  $135/70$  &  {\tiny Not rot.}\\
    CDFS-13497 & $3133/91$   &  {\tiny Not rot.}\\
    CDFS-14411 & $62.7/68$   &  Rotating\\
    CDFS-16272 & $48.4/82$   &  Not clas.\\ 
    CDFS-16767 &$63.0/46$   &  Rotating\\
    \hline
    \noalign{\smallskip}
    LSD Objects  &  & \\
    \hline
    \noalign{\smallskip}
    SSA22b-C5 & $216/138$   &  {\tiny Not rot.}\\
    SSA22a-C6 & $101/54$   &  {\tiny Not rot.}\\
    SSA22a-M4 & $21.8/36$   &  Not clas.\\
    SSA22a-C30 & $19334/243$   &  {\tiny Not rot.}\\
    Q0302-C131 & $62.2/95$  & Rotating\\
    Q0302-C171 & $170/102$   &  Not clas.\\
    DSF2237b-D28 & $1848/203$   &  {\tiny Not rot.}\\
    Q0302-M80 &  $206/87$  &  {\tiny Not rot.}\\
    DSF2237b-MD19 & $268/95$   &  {\tiny Not rot.}\\
    \hline
  \end{tabular} 
\end{table}

\section{Kinematical modeling}\label{s4}

\subsection{The model}\label{s41}
The adopted dynamical model assumes that the ionized gas is circularly rotating in a thin disk.
We neglect all hydrodynamical effects, therefore the disk motion is entirely determined by the gravitational potential.

In principle turbulent motions, often observed in high z rotating disk galaxies as well as in our own data, could provide significant dynamical support. Indeed \cite{Epinat:2009} apply an asymmetric drift correction assuming that turbulent motions support mass. However, while stellar chaotic motions always support mass because they have gravitational origin, this is not always true for gas motions. The gas velocity dispersion might be strongly increased because of non gravitational motions (like, e.g., in winds driven by starburst activity), and in that case it would not be directly linked to the dynamical mass. Therefore we adopted a conservative approach and neglected the contribution of the gas velocity dispersion. In any case we will, at most, underestimate the dynamical mass by a factor lower than $\sim2$ at the disk scale length \citep{Epinat:2009}.

The galaxy gravitational potential is generated by an exponential disk mass distribution with surface density given by:
\begin{equation}
\Sigma(r)=\Sigma_0\ e^{-r/r_e}
\end{equation} 
The rotation curve of a thin disk with due to such mass distribution is \citep{Binney:1987}:
\begin{equation}
V_c(r)^2=4\pi G\, \Sigma_0 r_e y^2 [I_0(y)K_0(y)-I_1(y)K_1(y)]
\end{equation}
where $y=r/2r_e$ and $I_n$, $K_n$ are the modified Bessel functions of the first and second kind.

By defining the ``dynamical mass'' of the galaxy as the total mass enclosed in a $10$ kpc radius (e.g.~\citealt{Cresci:2009}), we can finally write 
\begin{equation}
\label{1}
V_c(r)=6.56\mathrm{\,km\ s^{-1}}\,\left(\frac{M_{dyn}}{10^{10}M_{\sun}}\right)^{0.5}\left(\frac{r}{kpc}\right) A(y)
\end{equation}
where
\begin{equation}
\label{1}
A(y)=\left[\frac{I_0(y)K_0(y)-I_1(y)K_1(y)}{I_0(y_{10})K_0(y_{10})-I_1(y_{10})K_1(y_{10})}\right]^{0.5}
\end{equation}
and  $y_{10}$ corresponds to $y$ computed for $r = 10$ kpc.

The velocity along the line of sight for a given position on the sky is derived from $V_c(r)$ by taking into account geometrical projection effects.

Another important element of the model is the intrinsic flux distribution (hereafter IFD) of
the emission line, 
since it acts as a weighting function in the calculation of the observed velocities and velocity dispersions when taking into account instrumental effects.
Following \cite{Cresci:2009} the IFD in the disk plane is modeled with an exponential function.
\begin{equation}
I(r)=I_0 e^{-r/r_0}
\end{equation}
where $r$ is the distance from the disk center and $r_0$ is the scale radius. Such IFD is then projected onto the plane of the sky.
Several objects in our sample have surface brightness distributions clearly different from those of exponential disks. However, given the low S/N and poor spatial resolution of our data, we decided to keep the IFD modeling as simple as possible (see also \citealt{Cresci:2009}).
 
In computing the observed velocity shift and velocity dispersion we take into account the instrumental beam smearing. The spatial PSF is modeled with a two-dimensional gaussian function with full width half maximum $FWHM_{PSF}$, sampled over the detector $x$, $y$ pixels. The instrumental spectral response is also described by a Gaussian function with sigma $\sigma_{inst}$. 
Finally, we calculate the model values for the three observed quantities: $F_{model}(x_i,y_j)$, $v_{model}(x_i,y_j)$ and $\sigma_{model}(x_i,y_j)$ where $x_i, y_j$ represent a spatial pixel in the data cube.
For a detailed description of the model refer to appendix B of \cite{marconi:2003a}.

Summarizing, the model parameters are:
\[
  \begin{array}{lp{0.8\linewidth}}
     x_c, y_c      & coordinates in the plane of the sky of the disk dynamical centre.\\
     \theta      & position angle (PA) of the disk line of nodes.\\
     i            & inclination of the disk. \\
     M_{dyn}           &  dynamical mass. \\
     r_e           & characteristic radius of the exponential disk. \\
     V_{sys}      & systemic velocity of the galaxy.\\
  \end{array}
\]

By using these parameters values we derive $V_{max}$ that is the maximum velocity of the rotation curve. We will use this parameter in Sect.~\ref{s53} to build the Tully-Fisher relation for our data sample.

\subsection{Fit strategy}\label{s42}

\subsubsection{Preliminary steps}\label{s421}

The first preliminary step of the model fitting is to estimate the spectral and spatial resolution of the observations, characterized by the corresponding $\sigma_{spec}$ and $FWHM_{PSF}$ of the gaussian broadening functions.

The spectral resolution ($\sigma_{spec}$) is simply estimated by using the profile of isolated lines in sky spectra. 

Estimating the spatial resolution ($FWHM_{PSF}$) is more complex since we miss datacubes of stars obtained with the same instrumental setting at the same  time and at a similar position
on the sky as the object datacube. We have therefore devised a method to estimate $FWHM_{PSF}$ directly from the datacubes of the objects showing rotating-disk kinematics.

\begin{figure}[!ht]
  \centering
  \includegraphics[width=0.99\linewidth]{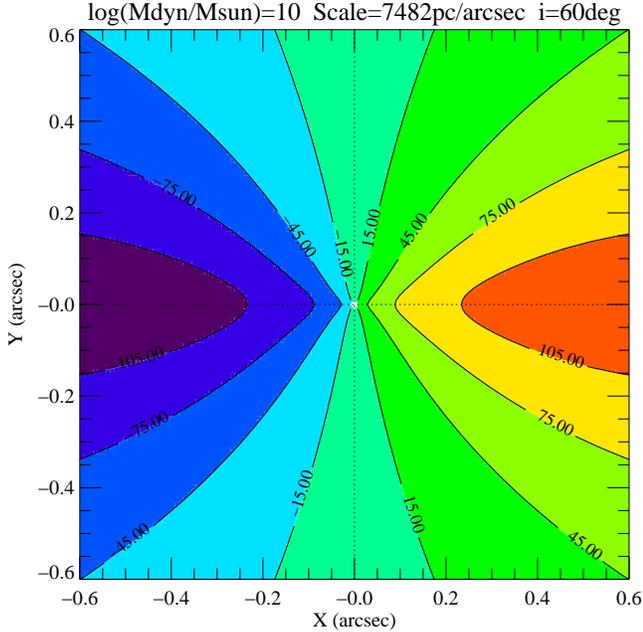}
  \caption{Velocity field on the sky plane (``spider diagram'') for a simulated rotating disk at redshift $z\sim3.3$ with mass of $10^{10}M_{\sun}$, an exponential mass profile with characteristic radius of $\sim 2.2kpc$, inclination of $60^{\circ}$ and velocities binned in steps of $30km\ s^{-1}$.}
  \label{fig02}
  \end{figure}

In the assumption that the kinematics is well approximated by a rotating disk we expect an intrinsic line of sight velocity field on the sky plane as shown in Fig.~\ref{fig02}, which
illustrates the iso-velocity contours projected on the sky of a simulated galaxy with mass $10^{10}M_{\sun}$ (the so called ``spider diagram''). The gas with line of sight velocity in a given velocity bin lies in the locus delimited by two subsequent isovelocity contours. We can observe that the spatial extent
of such region is always lower along the line of nodes direction (the $X$-axis) than in the perpendicular direction, and the central velocity bins show the minimal extent. For this reason
we expect that an image of the object in the central velocity bins will be spatially unresolved along the line of nodes direction (see the example of Fig. \ref{fig02}: the central velocity
bin of $30km\ s^{-1}$ has an extent along the line of nodes direction lower than $\sim0.1\arcsec$, well below the typical value of our spatial resolution of $\sim0.6\arcsec FWHM$).

In practice, we select the spatial planes of the object datacube with wavelength close the line centroid, and for each of these spatial planes we fit the line surface brightness with a two dimensional gaussian function. According to what is stated above, we take the minimum FWHM value from these two dimensional gaussians as an estimate of $FWHM_{PSF}$.

We could test our method with one object, {\it CDFa-C9}, for which there is a star in the SINFONI field of view. The FWHM of this star is consistent within $\sim0.1\arcsec$ with our estimate of the spatial resolution. We also verified that our estimates are in good agreement with the FWHM of the telluric standard stars observed after the science exposures, although with different airmasses. Moreover, in the following, we will test how systematic errors on $FWHM_{PSF}$ can affect the best fit model parameters.

The second preliminary step is the determination of the intrinsic flux distribution (IFD) to be used in the computation of the velocity maps.
The exponential IFD in the disk plane is projected onto the plane of the sky, convolved by the
PSF, averaged over the detector binsize and matched to the observed flux map by minimizing the following $\chi^2$
\begin{equation}
\chi^2_{flux}=\left[\sum_{i,j}{\frac{F(x_i,y_j)-F_{model}(x_i,y_j)}{\Delta F(x_i,y_j)}}\right]^2
\end{equation}

the free parameters in this fit are the center position of the IDF in the plane of the sky ($x_c$,$y_c$), the scale radius $r_0$ and the position angle and inclination of the elliptical isophotes ($I_0$,  the flux scale, is only a normalization factor).

From the fit of the IFD we also estimate some parameters needed in the model. As explained in the following section, we set the disk center position equal to the IFD center position ($x_c$,$y_c$) and the mass distribution scale radius $r_e$ equal to the IFD scale radius $r_0$. The inclination and position angle of the disk are instead determined from the fit of the velocity maps, but we use  the estimates from the IFD fit as first guesses.

\subsubsection{Fitting of velocity maps}\label{s422}

To estimate the free model parameters we minimize the following $\chi^2$:
\begin{equation}
\chi^2_{vel}=\left[\sum_{i,j}{\frac{v(x_i,y_j)-v_{model}(x_i,y_j)}{\Delta v(x_i,y_j)}}\right]^2
\end{equation}
In principle, all model parameters (disk center position, orientation and inclination, systemic
velocity and dynamical mass) should be left free to vary. However, the moderate S/N of the data
combined with the limited spatial coverage suggest to adopt a different approach in which some
of  these parameters are constrained a priori.

The mass distribution should be traced by the continuum flux distribution, therefore the disk center position should be identified by the continuum peak.
Only in some objects we can detect continuum emission with high enough S/N to locate the disk center. In such cases we use that position in the fit allowing a variation of  $\pm 0.1\arcsec$ consistent with measurement errors. For all other objects we use the position inferred from the fitting of the emission line flux maps. For those objects where we detect a continuum component the average distance between line and continuum peaks is $\sim0.25\arcsec$. Therefore, when we do not detect any continuum, we allow for a variation of the disk center position of $\pm 0.25\arcsec$.

The scale radius of the mass distribution,  $r_e$, cannot be derived from the velocity maps. Following \cite{Cresci:2009}, we fix $r_e$ to the value of $r_0$, the scaling radius of the continuum flux map. When this is not detected, we adopt the scale radius of the emission line flux map. Estimating the scale radius $r_e$ by fitting the IFD with a particular model (i.e. exponential) can of course lead to systematic errors. Therefore we will analyze for every object how systematic errors on $r_e$ affect dynamical mass values.

Another potential source of systematic errors is the use of the line flux map to estimate disk center position and scale radius when no continuum is detected. Therefore we also evaluated how this choice affects the dynamical mass: for the objects where the continuum is detected we performed the fit of the velocity maps using the disk center position and scale radii estimated either from continuum and line flux maps. We find an average variation of $0.2$~dex for the best fit $M_{dyn}$ values. This can be considered an estimate of the systematic error associated to the use of the line flux map instead of the continuum one.

Since the disk inclination is coupled with the total dynamical mass, we decide to keep it as a fixed parameter in the fitting procedure to avoid convergency problems with the $\chi^2$ minimization. Its best value and confidence interval is then identified 
by using a grid of $i$ values and performing the fit for each of them. In this way we construct the $\chi^2$ curve as a function of $i$ which allows us to identify both the best $i$ value and the confidence intervals (for details see \citealt{Avni:1976}).
In the cases where the inclination is not constrained by the fit, we set its value to $60^{\circ}$, which represents the mean value for a population of uniformly randomly oriented disks.

\subsubsection{The intrinsic velocity dispersion}\label{s423}

Our model computes the observed line velocity dispersion by taking into account the unresolved
rotation, the intrinsic instrumental broadening ($\sigma_{spec}$) and the broadening due to the
finite size of the spatial pixels and of the spatial PSF. However, the disk can be characterized by an intrinsic velocity dispersion due to turbulent and non-gravitational motions in general. We estimate such intrinsic velocity dispersion of the gas as 
\begin{equation}
\sigma_{int}(x_i, y_j)^2=\mathrm{max}[\,(\sigma_{observed}(x_i, y_j)^2- \sigma_{model}(x_i, y_j)^2)\,; 0]
\end{equation}
where $\sigma_{observed}(x_i, y_j)$ is the observed map of velocity dispersion and $\sigma_{model}(x_i, y_j)$ is that computed with the model which best fits the velocity map.

\subsubsection{Error estimates}\label{s424}
To estimate the errors on the best fitting model parameters we use the bootstrap method \citep{efron:1994}.
The  dataset of each galaxy consists of $n$ data points each characterized by spatial position,
flux, velocity and velocity dispersion.  We randomly extract from each dataset a subsample of
$n$ data points. Due to the random extraction, each subsample will have some data points that
are replicated a few times and some data points that are entirely missing.
We perform the fit on 100 subsamples and we then estimate errors on parameters by taking the standard deviation of the best fit values which are usually normally distributed.

\subsection{Fit Results}\label{s43}

The fit procedure outlined in the previous section is applied to the subsample of ``rotating'' objects in Tab.~\ref{tab2}.
In the following we present and discuss the results of the fit for each of these
objects. The observed, model and residual velocity maps are presented in the top panels of Figs.~\ref{fig03}-\ref{fig12}. The bottom panels of the same figures show the observed and model and intrinsic velocity dispersion maps.
The best fit parameters are presented in Tab.~\ref{tab3}.

\begin{description}

\item[\textbf{SSA22a-M38} (Fig. \ref{fig03}).]
  \begin{figure*}[!ht]
  \centering
  \includegraphics[width=0.75\linewidth]{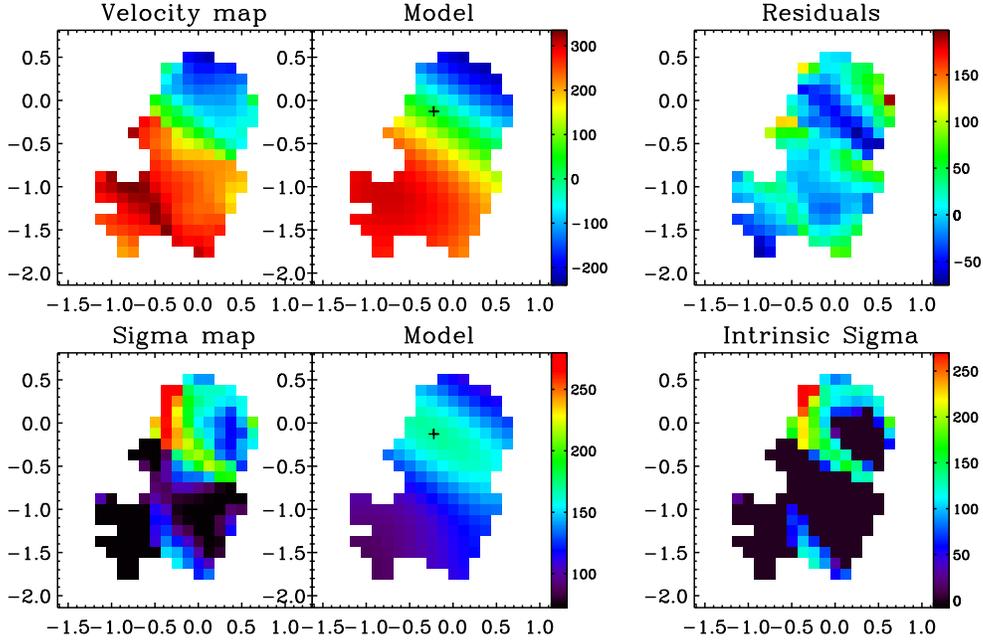}
  \caption{Fit results for the object {\it SSA22A-M38}. Respectively on top panels from the left: data, model and residuals map. On bottom panels from the left: data, model and ``intrinsic sigma'' map. The plus sign on the middle panels represents the position of the model dynamical center. The $X-Y$ coordinates are in arcsec referred to an arbitrary object center position. The North direction is the positive y axis. The vertical color bars are in $km\ s^{-1}$.}
  \label{fig03}
  \end{figure*}
In this object, as shown in Fig.~\ref{figa1}, we detect continuum emission which we use to estimate the position of the disk center.
The inclination is well constrained by the observations to $i=83^{\circ}\ ^{+2}_{-1}$ at
the $1\sigma$ confidence level indicating that the disk is seen edge-on.
This is a clear consequence of the fact that the observed isovelocity contours do not show any curvature as expected when the rotating disk is seen at intermediate inclinations. The best fitting dynamical mass is then little affected by systematic errors due to uncertainties on the disk inclination.
The FWHM of the spatial PSF estimated as described in Sec.~\ref{s421} is $0.5\arcsec$. To evaluate how
an incorrect estimate of this parameter can affect the $M_{dyn}$ value we repeated the fit
using FWHM in the range $0.4-0.6\arcsec$. The value of $M_{dyn}$ differ in this case by only
$0.07$~dex. The characteristic radius $r_e$ of each object depends on the value of the adopted PSF FWHM (e.g. setting a narrower PSF corresponds to an intrinsically broader flux distribution, with a larger $r_e$). Therefore we have also evaluated how an incorrect estimate of the scale radius $r_e$ affects the $M_{dyn}$ best fit value. We repeated the fit using $r_e$ different by $\pm30\%$ but keeping the PSF width fixed: the value of $M_{dyn}$ changes by only $\pm0.1$~dex.

By adding in quadrature all the systematic errors we obtain the values for the systematic errors on the dynamical masses reported in Tab.~\ref{tab3}.

In the bottom right panel of Fig. \ref{fig03} we present the intrinsic velocity dispersion that
is not accounted for by the ordered rotation of the gas. Most of the map is characterized by a
very low intrinsic dispersion with average value of $\langle \sigma_{int} \rangle
\sim50kms^{-1}$. This means that most of the rotating disk is dynamically "cold". However, a
North-East region is apparently characterized by high turbulence ($ \sigma_{int}
\sim250kms^{-1}$),
although it is not clear whether the large velocity dispersion at this location
is a consequence of low S/N of the emission lines.
There is the possibility that the weak clump observable in the flux map at the south of the
principal peak (Fig.~\ref{figa1}) could be a distinct object. In our analysis we assume that
all emission comes from an unique disk, but if the southern clump is really a distinct object we have to exclude
its emission from the disk. However our assumption is strengthened by the continuum flux map that do not show any
secondary clump, hence the secondary sourthern clump is likely a star forming region within the disk.

\item[{\bf SSA22a-C16} (Fig. \ref{fig04}).]
  \begin{figure*}[!ht]
  \centering
  \includegraphics[width=0.75\linewidth]{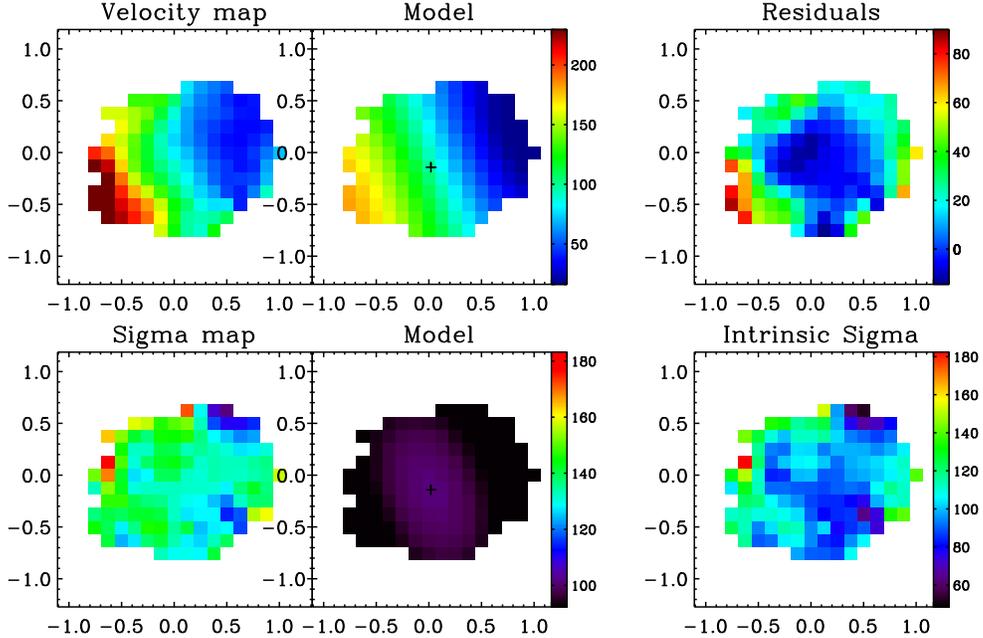}
  \caption{Fit results for the object {\it SSA22A-C16}. Respectively on top panels from the left: data, model and residuals map. On bottom panels from the left: data, model and ``intrinsic sigma'' map. Axis, symbols and color bars indication same as in Fig. \ref{fig03}.}
  \label{fig04}
  \end{figure*}
In this object we detect continuum emission that allows us to fix the center of rotation, but the S/N does not
allow us to estimate the scaling radius, which is then inferred from the flux distribution of the emission line.
The inclination is constrained to be $i=60^{\circ}\ ^{+15}_{-30}$ at the $1\sigma$ confidence level resulting in a systematic error on the mass of $+0.45\ -0.03$ dex.
The PSF width estimated is $FWHM_{PSF} =0.5\arcsec$. Considering a range of variation of $FWHM_{PSF}=0.4-0.6\arcsec$, the value of $M_{dyn}$ has a systematic error of $0.15$ dex. A variation of $30\%$ of the disk scale radius produces a variation of $0.1dex$ on the best fit $M_{dyn}$ value. By combining all the contributions to the systematic error we obtain a total error of $+0.48\ -0.18$ dex.
The intrinsic velocity dispersion maps in the bottom right panel of Fig. \ref{fig04} indicates that this object is turbulent, with a quite constant value of  $ \sigma_{int} \sim100kms^{-1}$.

\item[{\bf CDFS-2528} (Fig. \ref{fig05}).]
  \begin{figure*}[!ht]
  \centering
  \includegraphics[width=0.75\linewidth]{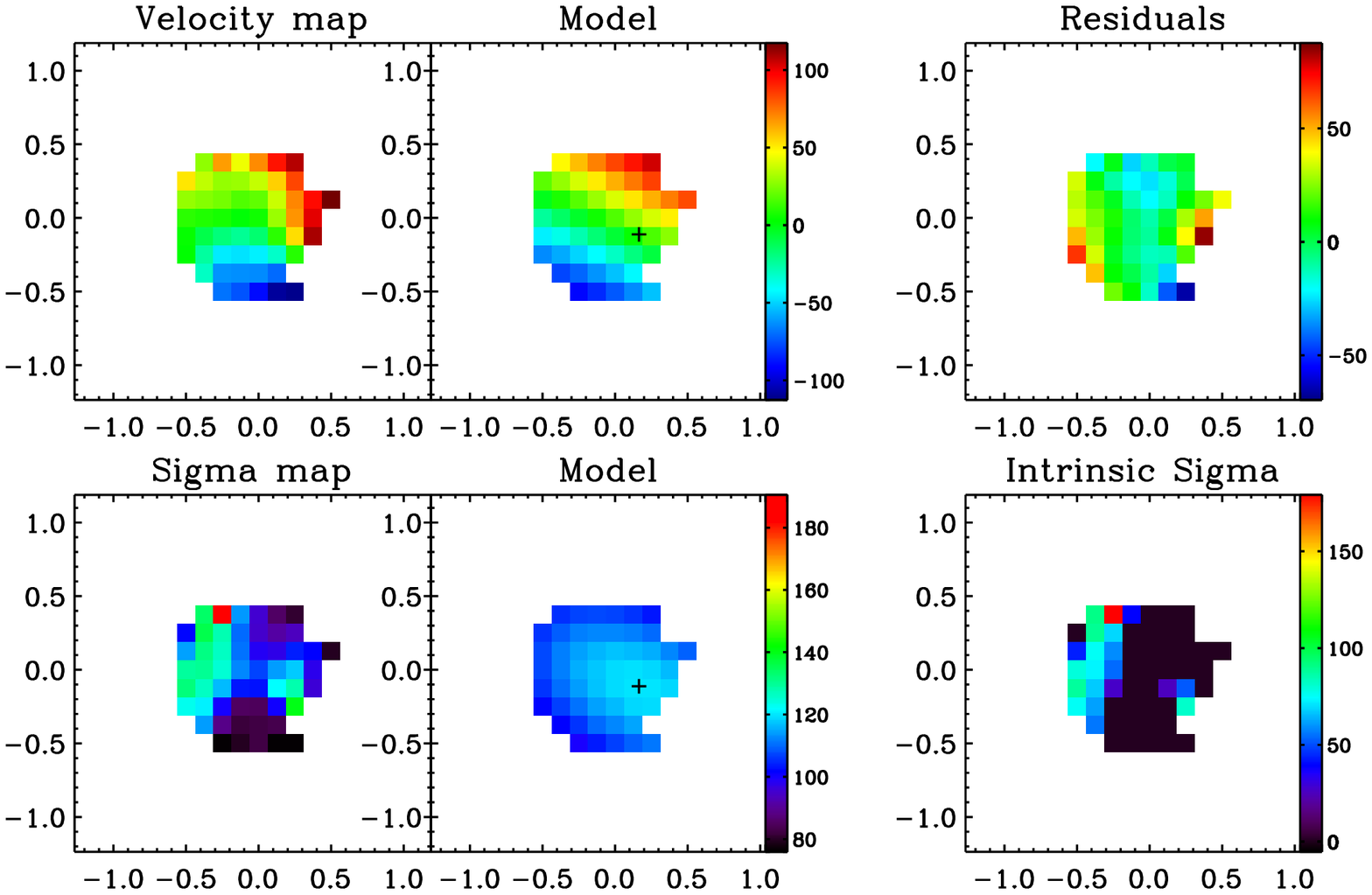}
  \caption{Fit results for the object {\it CDFS-2528}. Panels, axis, symbols and color bars indication same as in Fig. \ref{fig03}.}
  \label{fig05}
  \end{figure*}
The continuum emission detected in this object can only be used to constrain the position of  the center of rotation.
The inclination is only constrained to be  $i<70^{\circ}$, therefore we can only provide a lower limit on the
dynamical mass of $\log{(M_{dyn}/M_{\sun})}> 10.65 $. In this case, when inclination is partially or totally unconstrained, we estimate the mass by adopting the fiducial value of $i=60^\circ$.
The PSF estimated width is $FWHM_{PSF}=0.7\arcsec$. Allowing a variation in the range $FWHM_{PSF}=0.6-0.8\arcsec$ the value of $M_{dyn}$ varies by $0.07$ dex. A variation of $30\%$ of the disk scale radius produces a variation on $M_{dyn}$ of $0.1dex$. The intrinsic dispersion is quite low on average ($\langle \sigma_{int} \rangle \sim20 km s^{-1} $) with an excess of $\sim100kms^{-1}$ in the East part.

\item[{\bf SSA22a-D17} (Fig. \ref{fig06} ).]
  \begin{figure*}[!ht]
  \centering
  \includegraphics[width=0.75\linewidth]{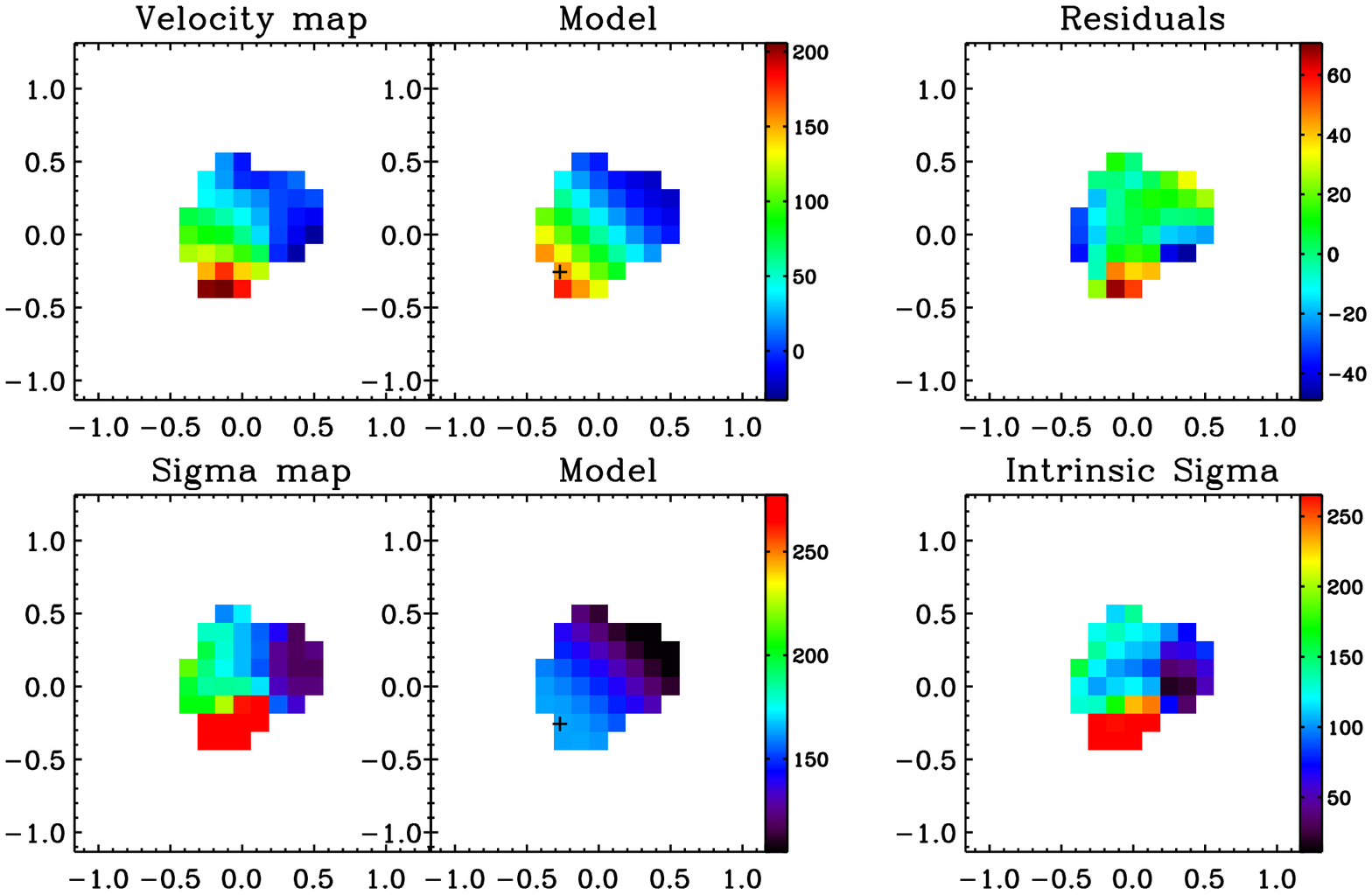}
  \caption{Fit results for the object {\it SSA22a-D17}. Panels, axis, symbols and color bars indication same as in Fig. \ref{fig03}.}
  \label{fig06}
  \end{figure*}
For this object we do not detect any continuum emission. The inclination is constrained to be  $i>70^{\circ}$ with a best fit value of $85^\circ$. The systematic error on the dynamical mass due to the inclination is then $+0.02\ -0.13$ dex.
The PSF width is $FWHM_{PSF}=0.65\arcsec$. Varying $FWHM_{PSF}$ by $\pm 0.1\arcsec$ we find a negligible systematic error on the dynamical mass.
  Varying by $30\%$ the disk scale radius produces a variation on $M_{dyn}$ of $0.08dex$. As previously noted, since we do not detect any continuum in this object we have to consider an additional systematic error of $0.2$~dex due to the use of the line flux distribution. The disk is quite turbulent with an average intrinsic dispersion of $\langle \sigma_{int} \rangle \sim130km s^{-1} $ and a southern peak of $\sim250kms^{-1}$.

\item[{\bf CDFa-C9} (Fig. \ref{fig07}).]
  \begin{figure*}[!ht]
  \centering
  \includegraphics[width=0.75\linewidth]{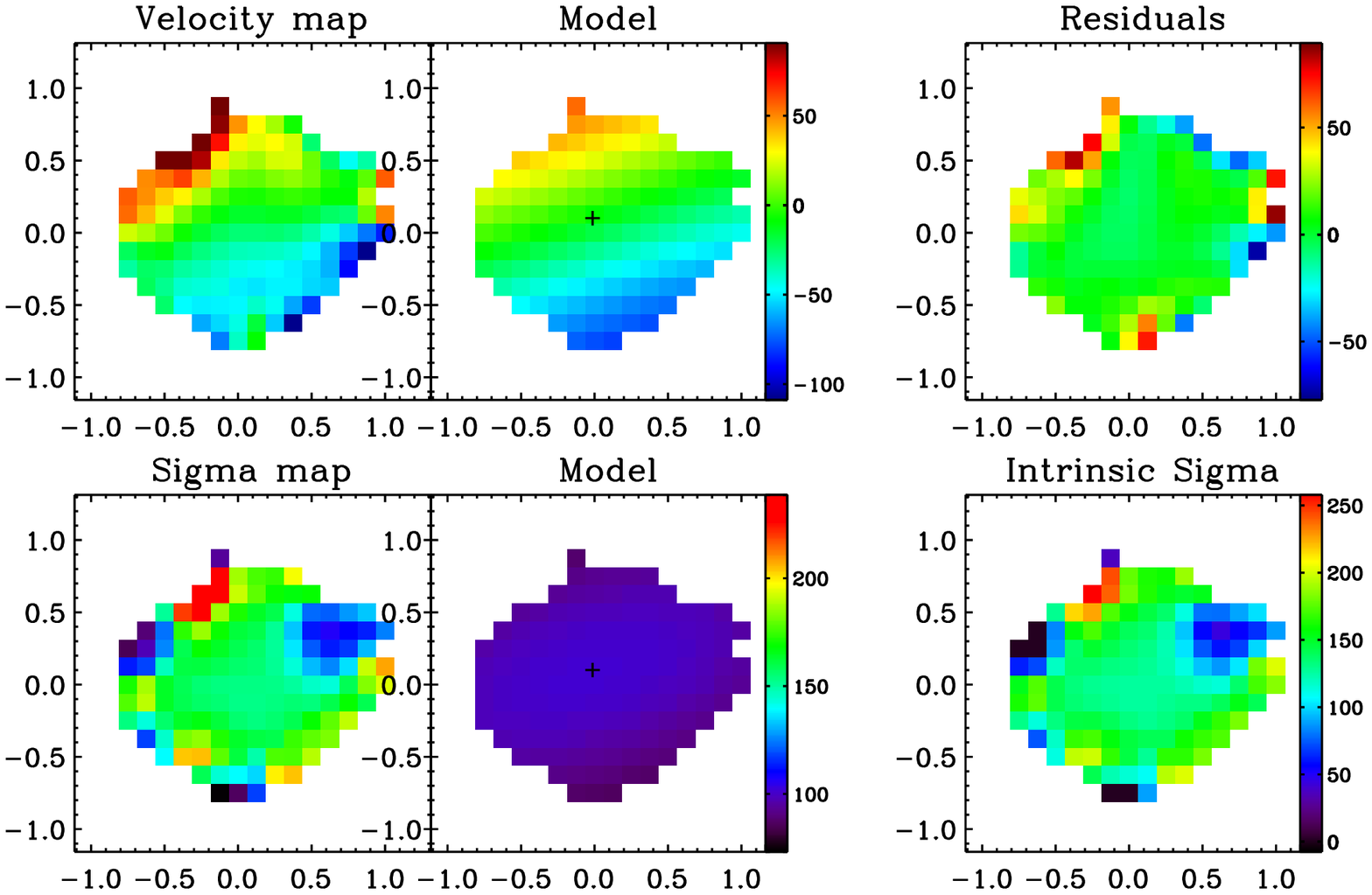}
  \caption{Fit results for the object {\it CDFa-C9}. Panels, axis, symbols and color bars indication same as in Fig. \ref{fig03}.}
  \label{fig07}
  \end{figure*}
For this object we detect the continuum, but we can only constrain the disk center position.  The inclination is constrained to be  $i>20^{\circ}$ with a best fit value of $60^\circ$. The systematic error on the dynamical mass due to the inclination is then $+0.8\ -0.1$ dex.
The PSF width is $FWHM_{PSF}=0.5\arcsec$. Varying $FWHM_{PSF}$ by $\pm 0.1\arcsec$ we find a systematic error on $M_{dyn}$ of $0.22$ dex.
  Varying by $30\%$ the disk scale radius produces a variation on $M_{dyn}$ of only $0.02dex$, therefore we can neglect this contribution to the systematic error. The disk is quite turbulent with an average intrinsic dispersion of $\langle \sigma_{int} \rangle \sim120km s^{-1}$.

\item[{\bf CDFS-9313} (Fig. \ref{fig08}).]
(In the same field as CDFS-9340).
The continuum is not detected.
  \begin{figure*}[!ht]
  \centering
  \includegraphics[width=0.75\linewidth]{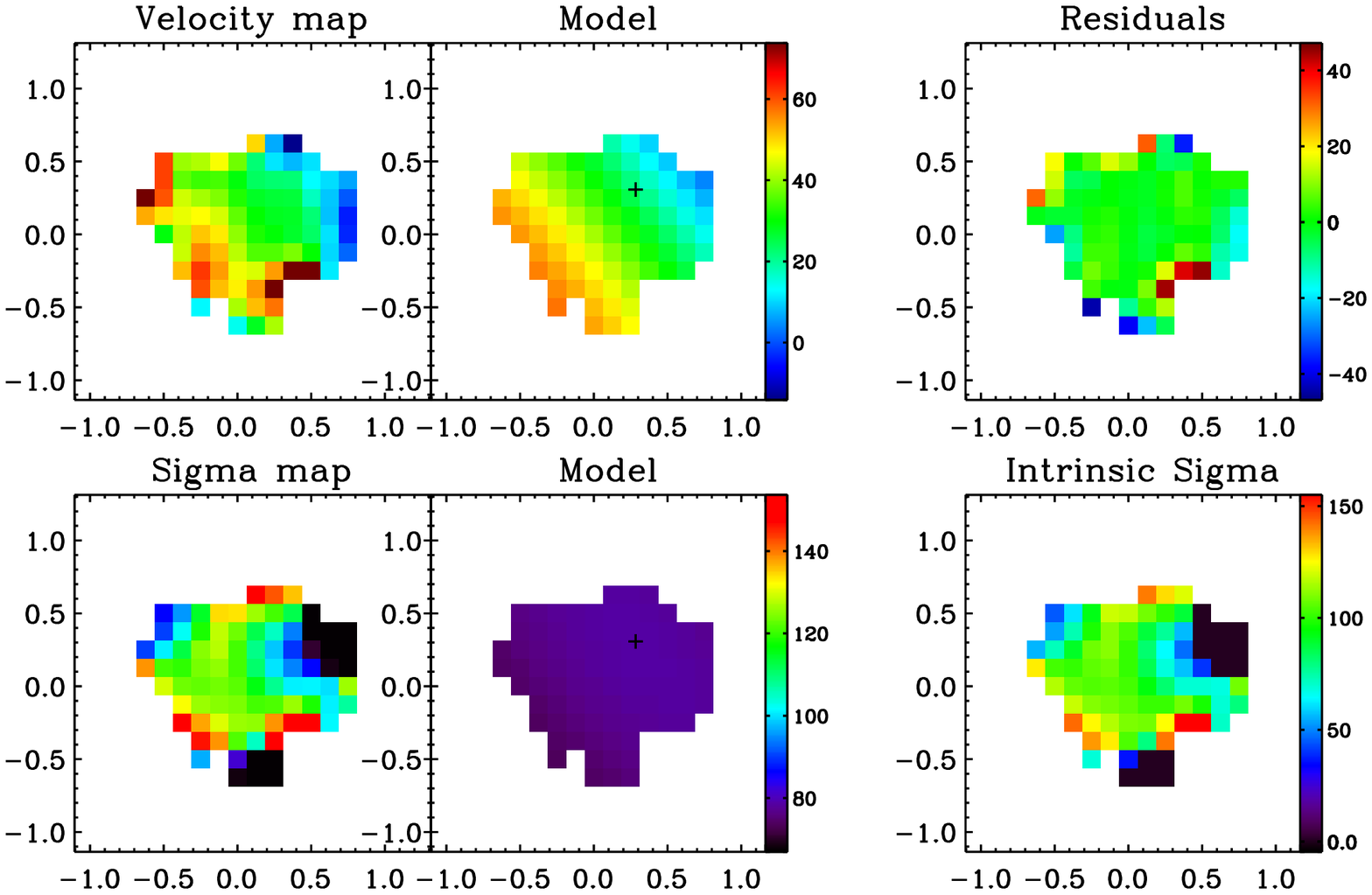}
  \caption{Fit results for the object {\it CDFS-9313}. Panels, axis, symbols and color bars indication same as in Fig. \ref{fig03}.}
  \label{fig08}
  \end{figure*}
We set the inclination to $i=60^{\circ}$ because it is totally unconstrained.
The PSF width is $FWHM_{PSF}=0.7\arcsec$. Varying $FWHM_{PSF}$ by $\pm 0.1\arcsec$ we find a $M_{dyn}$ systematic error of $0.88$ dex.
  Varying by $30\%$ the disk scale radius produces a variation on $M_{dyn}$ of only $0.02dex$, so we can neglect this contribution to the systematic error. The intrinsic velocity dispersion is quite constant over the entire map with an average value of $\langle \sigma_{int} \rangle \sim100km s^{-1}$.

  \begin{figure*}[!ht]
  \centering
  \includegraphics[width=0.75\linewidth]{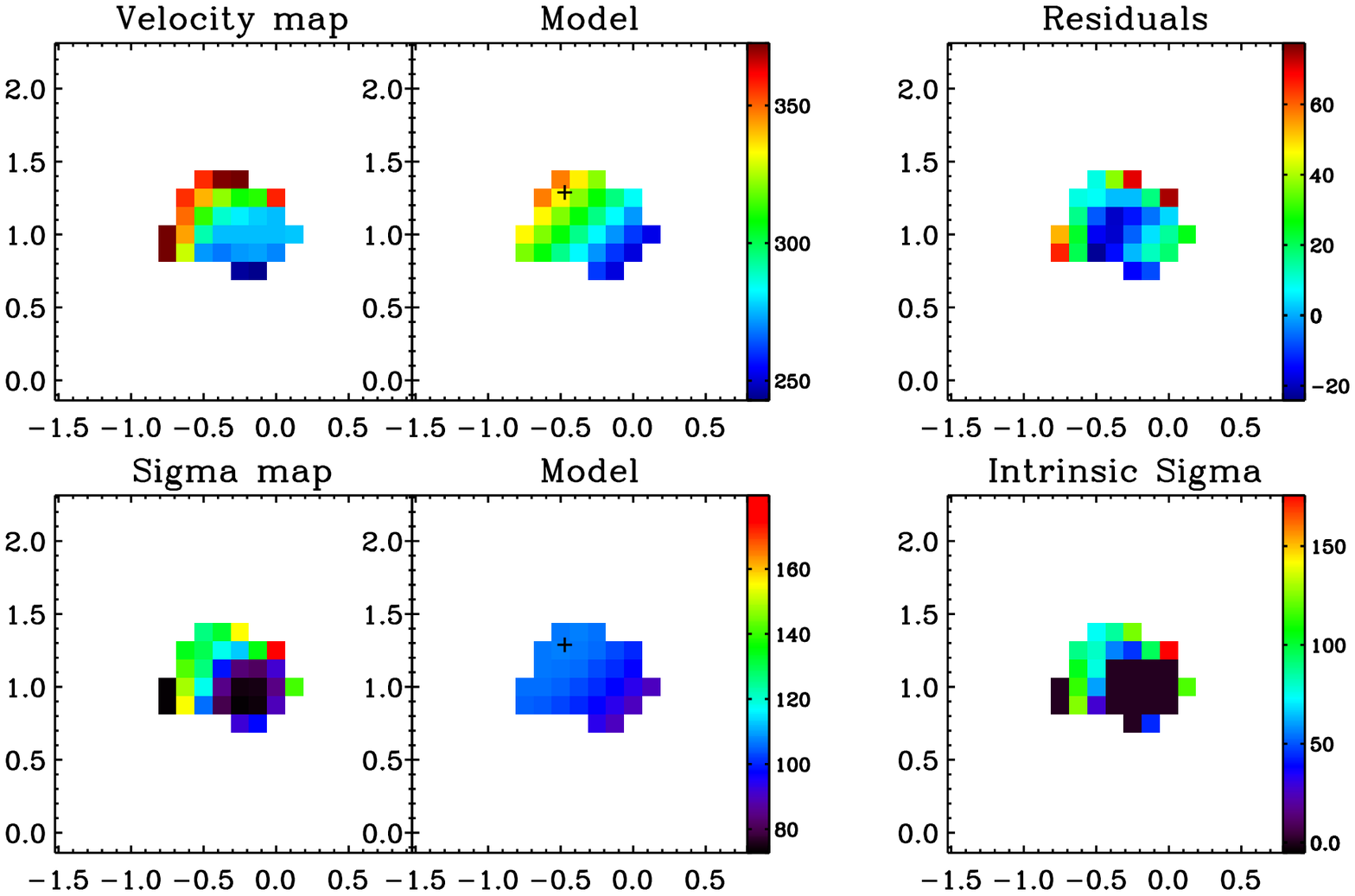}
  \caption{Fit results for the object {\it CDFS-9340}. Panels, axis, symbols and color bars indication same as in Fig. \ref{fig03}.}
  \label{fig08B}
  \end{figure*}
  
\item[{\bf CDFS-9340} (Fig. \ref{fig08B}).] (In the same field as CDFS-9313).
The continuum is not detected.
The inclination best fit value is $i=70^{\circ}$, but it is unconstrained at the $1\sigma$ confidence level.
The PSF width is the same estimated for {\it CDFS-9313}. Varying $FWHM_{PSF}$ by $\pm 0.1\arcsec$ we find a $M_{dyn}$ systematic error contribution of $0.1$ dex.  Varying by $30\%$ the disk scale radius produces a variation on $M_{dyn}$ of $0.1dex$. The intrinsic dispersion has an average value of $\langle \sigma_{int} \rangle \sim43kms^{-1}$.

The small separation of this object from {\it CDFS-9313} either in position ($\sim 1,0\arcsec-7.2kpc$) and in redshift ($\sim280kms^{-1}$) suggests to consider them as an interacting pair. The most massive object is in this case the least luminous, {\it CDFS-9340} (see tab.~\ref{tab3}). We estimate a mass ratio for the pair of $\sim10$, indicative of a minor merging event with a relatively large separation of the two objects (i.e. $\sim 7.2kpc$ compared to the disk scale radii of $0.7kpc$ and $1.4kpc$), consistently with the non perturbed rotating disk kinematics observed in the two objects.

\item[{\bf 3C324-C3} (Fig.~\ref{fig9}).]
  \begin{figure*}[!ht]
  \centering
  \includegraphics[width=0.75\linewidth]{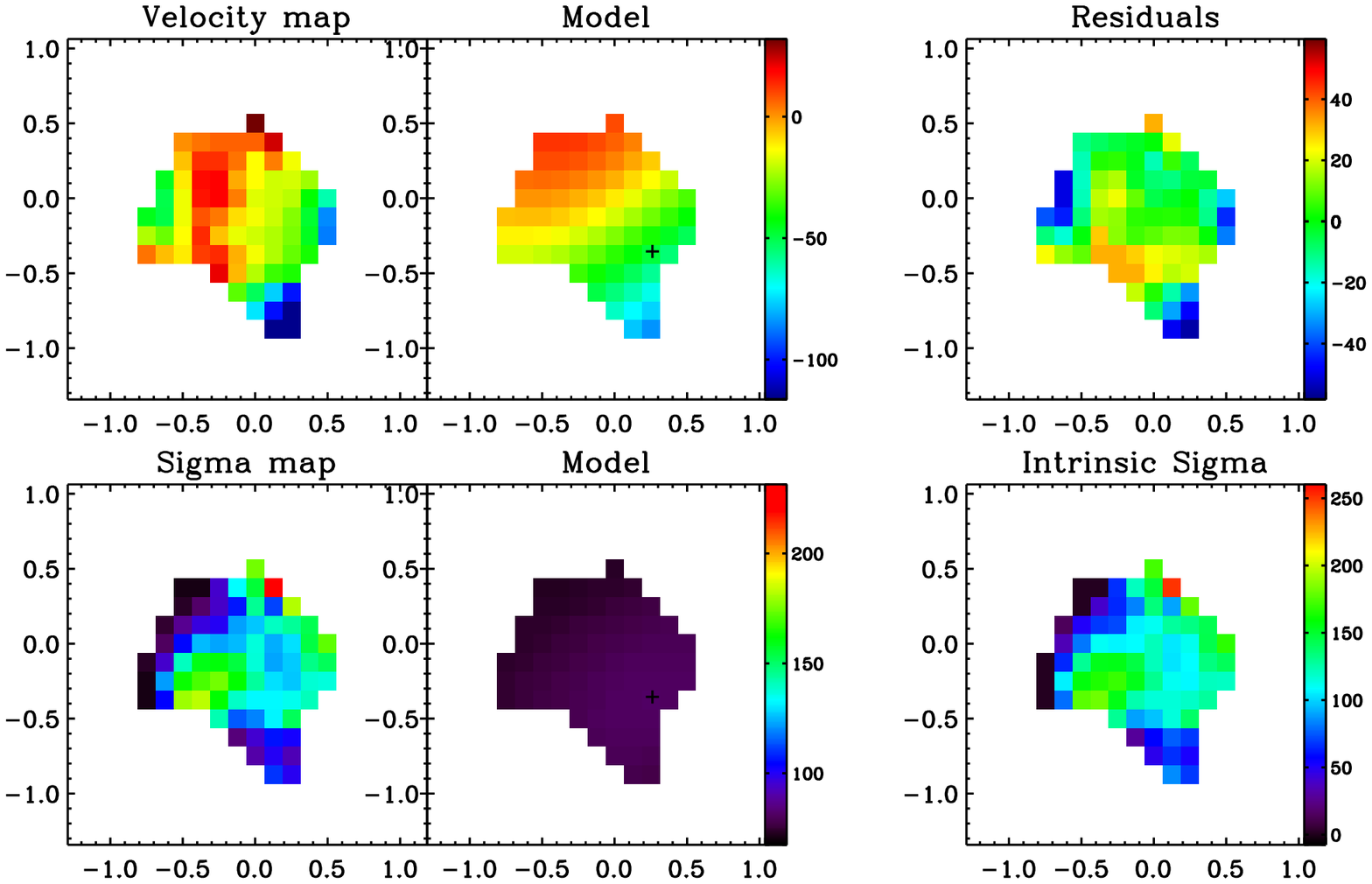}
  \caption{Fit results for the object {\it 3C324-C3}. Panels, axis, symbols and color bars indication same as in Fig. \ref{fig03}.}
  \label{fig9}
  \end{figure*}
In this object we do not detect any continuum emission. The inclination is constrained to be  $30^{\circ}<i<70^{\circ}$ with a best fit value of $60^\circ$ providing a systematic error on the dynamical mass of $+0.43\ -0.05$ dex.
The PSF width is $FWHM_{PSF}=0.7\arcsec$. Varying $FWHM_{PSF}$ by $\pm 0.1\arcsec$ we find a negligible $M_{dyn}$ systematic error contribution ($0.001$  dex). Varying by $30\%$ the disk scale radius results in a variation of $M_{dyn}$ by $0.1dex$. We note that the observed velocity field although showing a regular gradient is quite different from what expected in the case of a rotating disk. The intrinsic velocity dispersion is quite constant over the entire map with an average value of $\langle \sigma_{int} \rangle \sim125km s^{-1}$.

\item[{\bf CDFS-14411} (Fig. \ref{fig10}).]
  \begin{figure*}[!ht]
  \centering
  \includegraphics[width=0.75\linewidth]{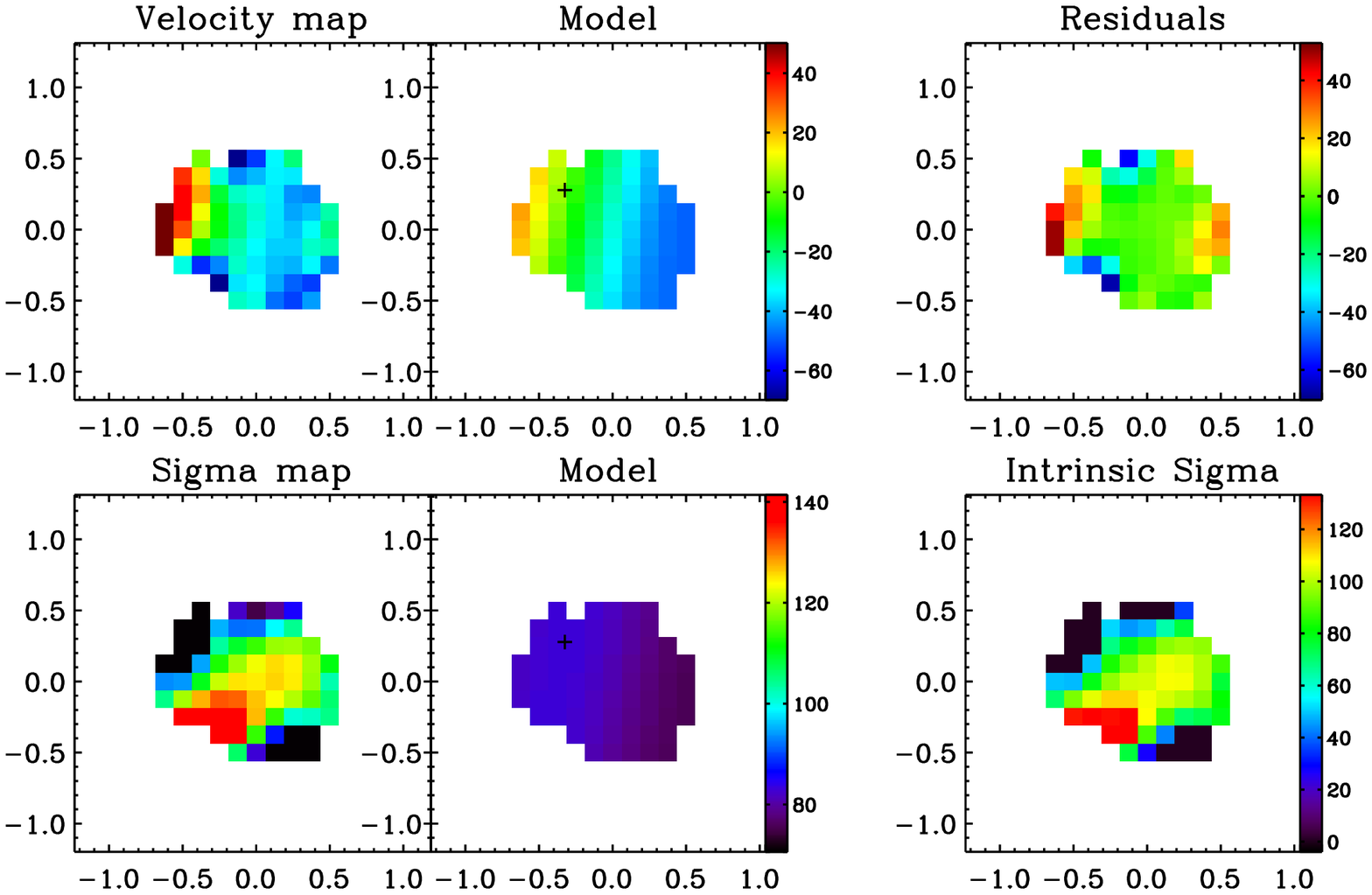}
  \caption{Fit results for the object {\it CDFS-14411}. Panels, axis, symbols and color bars indication same as in Fig. \ref{fig03}.}
  \label{fig10}
  \end{figure*}
For this object we do detect a continuum emission and use it to constrain the disk center position and scale radius. The inclination is constrained to be  $80^{\circ}<i<88^{\circ}$ with a best fit value of $85^\circ$. 
The dynamic mass systematic error due to the inclination is then negligible ($\pm 0.04$ dex).
The PSF width is $FWHM_{PSF}=0.6\arcsec$. Varying $FWHM_{PSF}$ by $\pm 0.1\arcsec$ we find a $M_{dyn}$ systematic error contribution of $0.17$~dex. Varying by $30\%$ the disk scale radius varies $M_{dyn}$ by $0.12dex$. The intrinsic velocity dispersion rises in the center ($\sigma_{int}\sim110kms^{-1}$). The mean value over the entire map is $\langle \sigma_{int} \rangle \sim65km s^{-1} $.

\item[{\bf CDFS-16767} (Fig.~\ref{fig11}).]
  \begin{figure*}[!ht]
  \centering
  \includegraphics[width=0.75\linewidth]{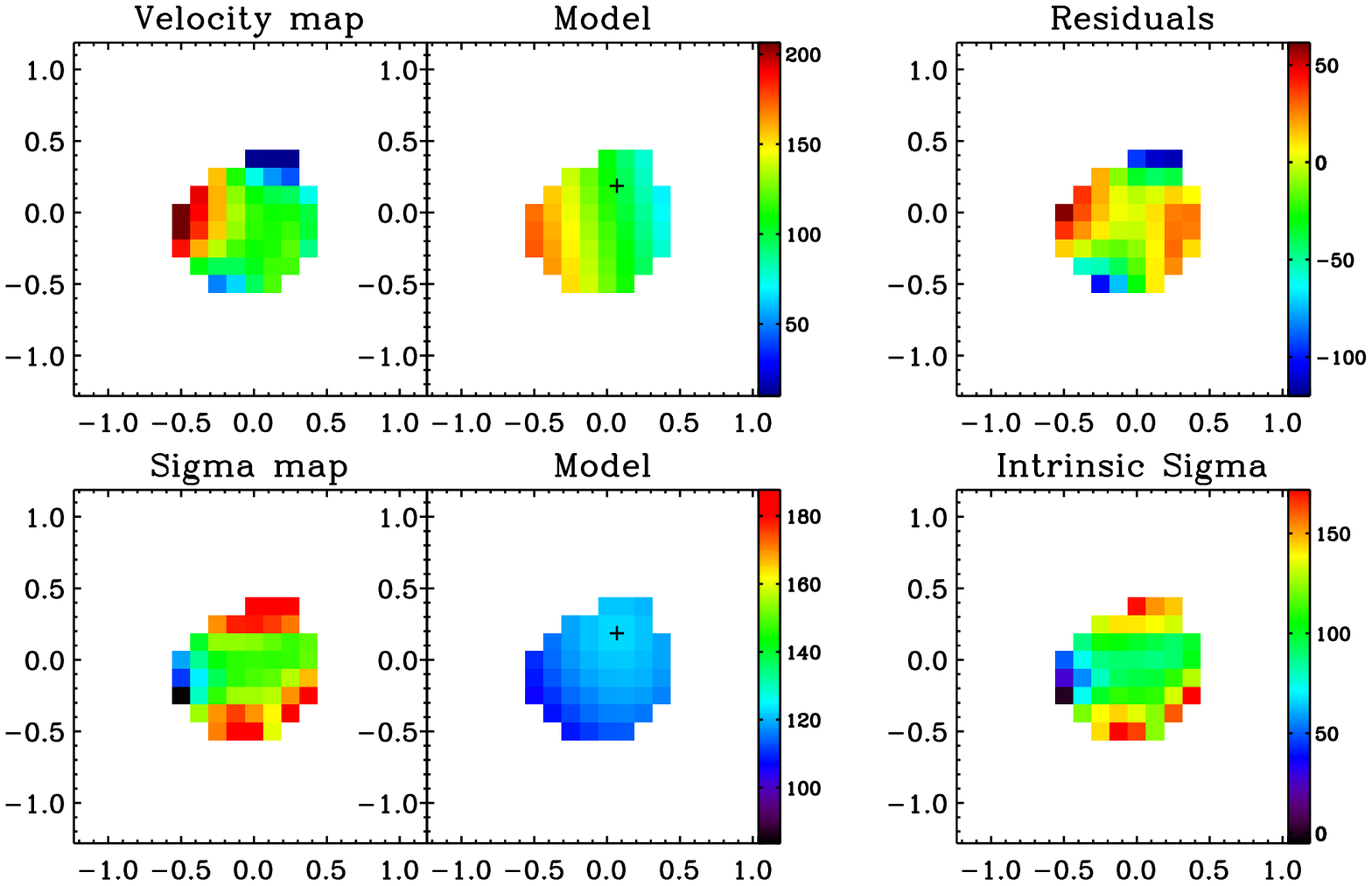}
  \caption{Fit results for the object {\it CDFS-16767}. Panels, axis, symbols and color bars indication same as in Fig. \ref{fig03}.}
  \label{fig11}
  \end{figure*}
In this object we detect the continuum emission and use it to constrain the disk center position and scale
radius. The inclination best value is $i=15^{\circ}$, but it is unconstrained at the $1\sigma$ confidence level.
The PSF width is $FWHM_{PSF}=0.55\arcsec$. Varying $FWHM_{PSF}$ by $\pm 0.1\arcsec$ we find a $M_{dyn}$ systematic error contribution of $0.21$~dex. Varying of $30\%$ the disk scale radius produces a negligible variation on $M_{dyn}$ ($0.04$~dex). The disk is characterized by high turbulence with an average intrinsic dispersion of $\langle \sigma_{int} \rangle \sim100km s^{-1} $.

\item[{\bf Q0302-C131} (Fig.~\ref{fig12}).]
  \begin{figure*}[!ht]
  \centering
  \includegraphics[width=0.75\linewidth]{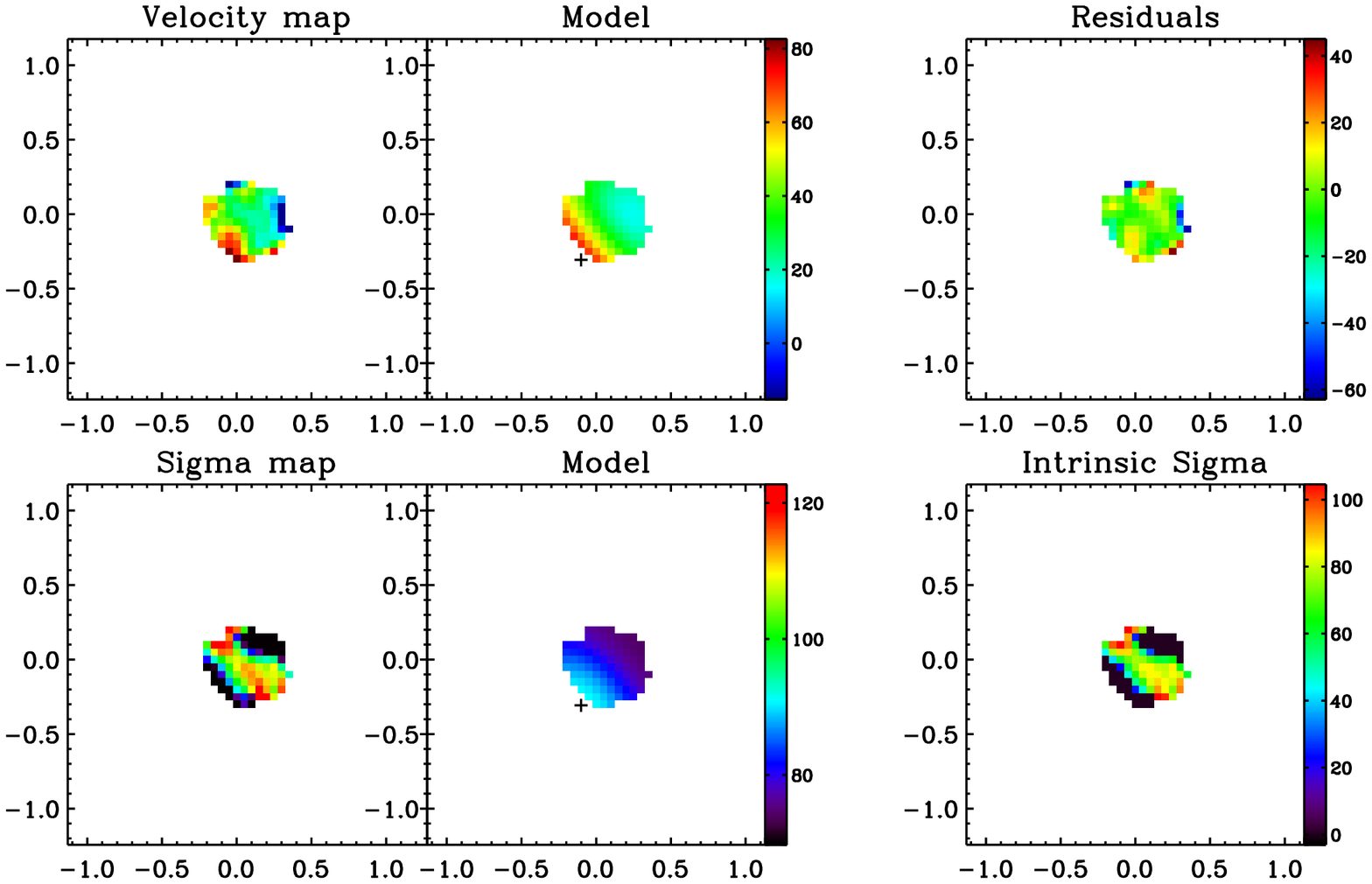}
  \caption{Fit results for the object {\it Q0302-C131}. Panels, axis, symbols and color bars indication same as in Fig. \ref{fig03}.}
  \label{fig12}
  \end{figure*}
This is the only rotating object in the LSD sample. In this object we do not detect any continuum emission. 
The inclination is constrained to be $i<60^{\circ}$, with a best fit value of $30^\circ$ therefore we can only provide
a lower limit on the dynamical mass of $\log{(M_{dyn}/M_{\sun})}>9.59$. The PSF width is $FWHM_{PSF}=0.3\arcsec$. Varying $FWHM_{PSF}$ by $\pm 0.1\arcsec$ we find a $M_{dyn}$ systematic error contribution of $0.88$~dex. Varying by $30\%$ the disk scale radius results in a negligible variation on $M_{dyn}$ ($0.01$~dex).
The disk has a quite low average intrinsic dispersion ($\langle \sigma_{int} \rangle \sim43km s^{-1} $).

\end{description}

\begin{table*}
  \caption[!ht]{Fit Results.}
  \label{tab3}
  \centering
  \begin{tabular}{l c c c c c c }
    \hline
    \hline
    \noalign{\smallskip}
     Object & $FWHM_{PSF}[\arcsec]$ & \multicolumn{5}{c}{Model parameter} \\
    \noalign{\smallskip}
    \hline
    \hline
    \noalign{\smallskip}
     &  & $\theta\ [^{\circ}]$ & $i\ [^{\circ}]\ ^{(1)}$ & $[log_{10}(M_{dyn}/M_{\sun})]$ & $r_e [kpc]\ ^{(2)}$ & $V_{max}[km\,s^{-1}]$ \\
    \noalign{\smallskip}
    \hline
    \noalign{\smallskip}
    SSA22a-M38 &  $0.5$ &  $149\pm2$ &  $83^{+2}_{-1}$ & $11.34\pm0.02\ (\pm0.12)^{(3)}$ & $2.574$ & $346^{+51} _{-45}$\\
    \noalign{\smallskip}
    SSA22a-C16 & $0.5$ &  $-111 \pm2$ &  $60^{+15}_{-30}$ & $10.31\pm0.02\ (^{+0.48}_{-0.18})^{(3)}$ & $1.730$ & $129^{+95}_{-24}$\\
    \noalign{\smallskip}
    CDFS-2528 & $0.7$ &  $-25\pm6$ &  $60\ (<70)$ & $10.71\pm0.05\ (>10.65)^{(4)}$ & $1.644$ & $>197^{(4)}$\\ 
    \noalign{\smallskip}
    SSA22a-D17 & $0.65$ &  $-45\pm5$ &  $80\ (>70)$ & $10.70\pm0.06\ (^{+0.22}_{-0.25})^{(3)}$ & $1.214$ & $260^{+75}_{-65}$ \\
    \noalign{\smallskip}
    CDFA-C9 & $0.5$ &  $18\pm2$ &  $60\ (>20)$ & $9.93\pm0.03\ (^{+0.83}_{-0.24})^{(3)}$ & $0.821$ & $129^{+206}_{-31}$ \\
    \noalign{\smallskip}
    CDFS-9313 & $0.7$ &  $128\pm4$ &  $60^{(5)}$ & $9.29\pm0.04\ ^{(5)}$ & $0.713$ & $67^{(5)}$ \\ 
    \noalign{\smallskip}
    CDFS-9340 & $0.7$ &  $44\pm12$ &  $70^{(5)}$ & $10.3\pm0.1\ ^{(5)}$ & $1.420$ & $151^{(5)}$ \\ 
    \noalign{\smallskip}
    3C324-C3 & $0.7$ &  $32\pm7$ &  $60^{+10}_{-30}$ & $10.12\pm0.12\ (^{+0.48}_{-0.23})^{(3)}$ & $2.5$ & $86^{+63}_{-20}$\\ 
    \noalign{\smallskip}
    CDFS-14411 & $0.6$ &  $78\pm5$ &  $85^{+3}_{-5}$ & $9.59\pm0.04\ (\pm0.21)^{(3)}$ & $1.240$ & $69^{+19}_{-15}$ \\
    \noalign{\smallskip}
    CDFS-16767 & $0.55$ &  $98\pm12$ &  $15^{(5)}$ & $10.93\pm0.11\ ^{(5)}$ & $0.362$ & $623^{(5)}$ \\ 
    \noalign{\smallskip}
    Q0302-C131 & $0.3$ &  $48\pm5$ &  $30\ (<60)$ & $9.99\pm0.07\ (>9.59)^{(4)}$ & $0.466$ & $>117^{(4)}$ \\ 
    \noalign{\smallskip}
    \hline
  \end{tabular} 
\begin{list}{}{}
\item[$(1)$] Parameter hold fixed, confidence interval estimated as explained in the text.
\item[$(2)$] Parameter hold fixed.
\item[$(3)$] Systematic error. Obtained combining the systematic error contributes due to the inclination estimate and to the PSF width estimate
\item[$(4)$] Upper-lower limit.
\item[$(5)$] Parameter unconstrained: we can not give a systematic error contribute due to this parameter.
\end{list}
\end{table*}

\section{Discussion}\label{s5}

\subsection{Rotating galaxies, turbulence and dynamical masses.}\label{s52}

In the AMAZE sample $\sim 40\%$ of the galaxies show a smooth velocity gradient (10 out of 24)
consistent with rotating disk kinematics. In the LSD sample only $\sim 13\%$ of the galaxies show a smooth velocity
gradient (1 out of 9). The lower fraction of rotating objects in the LSD sample is likely due to the
lower sensitivity of these observations to the outer, low surface brightness region
(which dominate the ``rotation signal'') because of the finer pixel scale, which makes the detector read-out noise more important relative
to the AMAZE observations performed with larger pixel scale. Moreover, the AMAZE sample also tend to have
a higher fraction of bright and redder targets (generally translating into higher masses), which may imply
a larger fraction of ``settled'' massive systems. We note that the fraction of rotating galaxies is fully
consistent, within statistical fluctuations, with the fraction of rotating galaxies ($\sim 33\%$)
found in the SINS sample at z$\sim$1.5--2.5 (\cite{Forster-Schreiber:2009}). This is particularly interesting,
given that the latter sample includes not only rest-frame UV selected galaxies, but also near- and mid-IR selected, and that
the fraction of rotating galaxies does not appear to depend on the selection criteria. Taken at face value,
the comparison between the SINS analysis at z$\sim$2 and the AMAZE analysis at z$\sim$3.3 suggests
that the fraction of rotating objects does not evolve within this redshift interval. \cite{Epinat:2009} derive a fraction of rotating galaxies of $22\%$ in a $z\sim1.2-1.6$ sample, but a direct comparison with this value is more difficult considering that their definition of rotating objects is quite different from the one adopted here.

In any case, these results should be taken with great caution since several observational effects may easily affect
the capability of identifying rotating objects. More specifically, the lack of angular resolution
may prevent the detection of a rotation pattern or, vice versa, may blend two merging systems mimicking
a rotation curve. Moreover, the limited sensitivity to low surface brightness disks (which is a strong
function of redshift) may prevent us to identify large rotating disks in distant galaxies.

In the following we compare the ordered rotation  (the velocity gradient $\Delta v$ over the map) with the random motions (the mean velocity dispersion).

  \begin{figure}[!ht]
  \centering
  \includegraphics[width=0.99\linewidth]{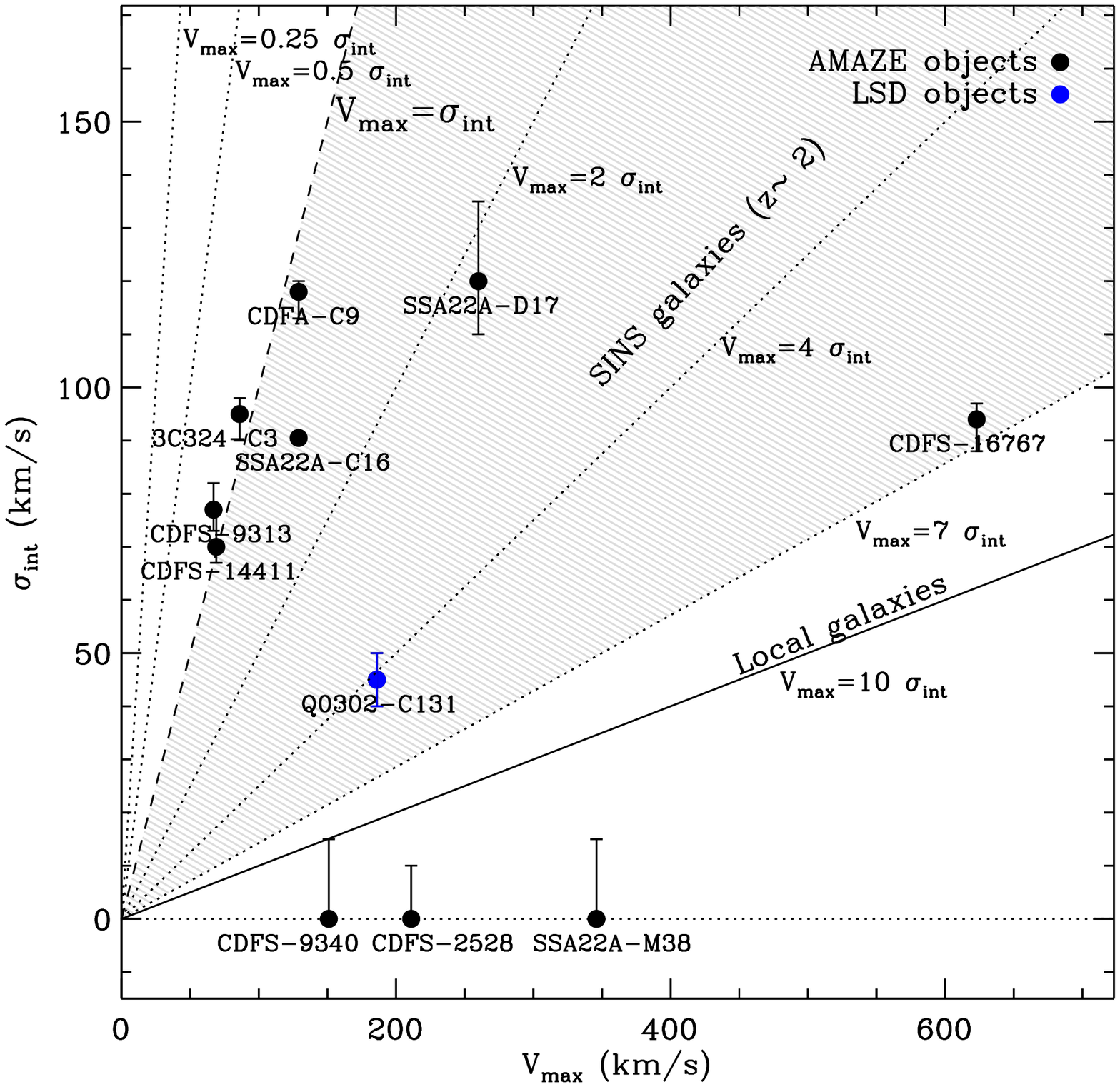}
  \caption{Maximum rotational velocity $V_{max}$ versus intrinsic velocity dispersion $\sigma_{int}$. Objects from
  the AMAZE and LSD samples are marked with black and blue symbols, respectively. The $\sigma_{int}$ value is obtained fitting the observed velocity dispersion map and error bars
  represent $1\,\sigma$ uncertainties. Overplotted the $V_{max}=\sigma_{int}$ locus (dashed line),  the $V_{max}=k\,\sigma_{int}$ loci (dotted lines), the mean value of $V_{max}/\sigma$ for the local galaxies (continuous line) and the $V_{max}/\sigma$ interval for the SINS sample at $z\sim2$ \citep{Forster-Schreiber:2009} (gray region).}
  \label{fig13}
  \end{figure}


By using the results of our dynamical modeling we obtain the maximum rotational velocity ($V_{max}$) and the intrinsic gas velocity dispersion
$\sigma_{int}$ (i.e. the turbulent gas motions). In Fig.~\ref{fig13} we plot $V_{max}$ versus $\sigma_{int}$ for our rotating objects. $\sigma_{int}$ is a constant value added in quadrature to the model velocity dispersion and determined, as in \cite{Cresci:2009}, by fitting the observed velocity dispersion map. The line $V_{max}=\sigma_{int}$ discriminate the loci of ``rotation dominated'' ($V_{max}  > \sigma_{int}$) and ``dispersion dominated'' objects ($V_{max} < \sigma_{int}$) and, for comparison, in local disk galaxies $V_{max} /\sigma\sim 10$.
From Fig.~\ref{fig13} it is clear that, with  $V_{max}  / \sigma_{int} < 2$, the majority of the rotating galaxies at z$\sim$3 (6 out of 11) is composed of dynamically ``hot'' disks, in contrast with local galaxies. Two galaxies have a larger $V_{max}/\sigma_{int}$ ratio but still smaller then local galaxies, while three galaxies are dynamically ``cold''. The finding that high-z disk galaxies are much more turbulent than local galaxies was already obtained by other studies at z$\sim$1-2 (e.g. \citealt{Genzel:2006}, \citealt{Forster-Schreiber:2006}) and suggestive
of high gas fractions making disks gravitationally unstable (\citealt{Tacconi:2010}, \citealt{Daddi:2010}). \cite{Forster-Schreiber:2009} obtain with the SINS sample $V_{max} /\sigma \sim 1-7$ with a mean value of $4.5$. Our result at z$\sim$3.3 ($\langle V_{max}/\sigma_{int}\rangle _{z=3.3} \sim 2$) goes in the same direction and is even more extreme considering that the majority of our galaxies have $V_{max}/\sigma_{int}$ comparable with the smallest values observed in the SINS survey. We do find however a few galaxies which are dynamically ``cold'' even when compared to local galaxies.The missing link between high redshift and local galaxies seems to be provided by \cite{Epinat:2009} who obtain $\langle V_{max}/\sigma_{int}\rangle \sim 3.5$ (at $z\sim1.2-1.6$) when considering the objects classified as ``rotating disks'' and ``perturbed rotators''. This number rises to $7.2$ when considering only the ``rotating disks''.

\begin{figure}[!ht]
  \centering
  \includegraphics[width=0.99\linewidth]{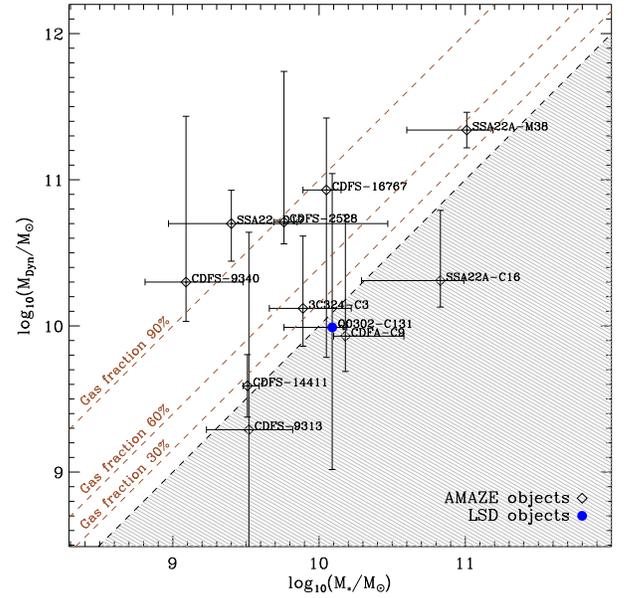}
  \caption{Comparison of the dynamical and stellar mass for the objects analyzed in this paper.
  The masked gray region represent the non physical condition of $M_{\star}>M_{dyn}$. The brown
  dashed lines represents the loci of constant gas fraction ($90\%$, $60\%$ and $30\%$ gas
  fraction, from left to right, respectively).}
  \label{fig14}
  \end{figure}

In Fig.~\ref{fig14} we compare dynamical and stellar masses for the rotating objects. We estimate dynamical masses in the range $\sim1.8\times10^9 M_{\sun}-2.2\times 10^{11} M_{\sun}$ with a mean value of $1.6\times10^{10} M_{\sun}$. The stellar masses are reported in table \ref{tab4} and are estimated from standard broad-band SED fitting (see Sect.~6 of \cite{Maiolino:2008} and Sect.~4.3 of \cite{Mannucci:2009} for more details).

\begin{table}
  \caption[!ht]{Stellar masses and Star formation rates.}
  \label{tab4}
  \centering
  \begin{tabular}{l c c}
    \hline
      \noalign{\smallskip}
     Object &  $[log_{10}(M_{\star}/M_{\sun})]$ & SFR $[M_{\sun}/y]$\\
    \noalign{\smallskip}
    \hline
    \noalign{\smallskip}
    SSA22a-M38 &  $11.01^{+0.18} _{-0.41}$&  $115^{+0.} _{-0.}$\\
    \noalign{\smallskip}
    SSA22a-C16 &  $10.83^{+0.16}_{-0.54}$&  $651^{+0.} _{-0.}$\\
    \noalign{\smallskip}
    CDFS-2528 & $9.76^{+0.09}_{-0.07}$&  $101^{+5.9} _{-53.8}$\\ 
    \noalign{\smallskip}
    SSA22a-D17 & $9.40^ {+1.07}_{-0.43}$&  $45.1^{+0.} _{-0.}$ \\
    \noalign{\smallskip}
    CDFA-C9 & $10.18^{+0.40}_{-0.08}$&  $265^{+0.} _{-0.}$ \\
    \noalign{\smallskip}
    CDFS-9313 & $9.52^{+0.29}_{-0.28}$&  $31.3^{+29.4} _{-23.9}$ \\ 
    \noalign{\smallskip}
    CDFS-9340 & $9.09^{+0.39}_{-0.28}$&  $11.7^{+11} _{-9.1}$ \\ 
    \noalign{\smallskip}
    3C324-C3 & $9.95^{+0.37}_{-0.18}$&  $88.6^{+68.8} _{-78.6}$ \\ 
    \noalign{\smallskip}
    CDFS-14111 & $9.51^{+0.08}_{-0.03}$&  $71^{+8.1} _{-3.8}$ \\
    \noalign{\smallskip}
    CDFS-16767 & $10.06^{+0.10}_{-0.16}$&  $84^{+89.1} _{-9.1}$ \\ 
    \noalign{\smallskip}
    Q0302-C131 & $10.09^{+0.10}_{-0.33}$&  $10^{+6} _{-4}$\\ 
    \noalign{\smallskip}
    \hline
  \end{tabular} 
\end{table}

All of the objects are consistent with having stellar masses lower than dynamical masses within the $1\sigma$ uncertainties.
By ascribing the difference between dynamical and stellar mass to the gas mass, the locii of constant
gas fraction are given by the dashed lines. Here we are assuming negligible contribution of Dark Matter,
justified by the fact that the spatial region sampled in our dynamical study is of the order of only $2$-$3$ characteristic radii
of the exponential disk, where the contribution of dark matter is still small. In any case, if this
assumption is not appropriate, we can call this quantity a ``gas and dark matter'' fraction.
Three objects require at high confidence a gas fraction of 90\% or higher. One object (SSA22a-C16)
has little room for any gas mass. Most of the other galaxies have the gas fraction poorly constrained or
totally unconstrained, except for SSA22a-M38 (with a gas fraction constrained within 30\% and 80\%)
and CDFS-14411 (with a gas fraction constrained to $<$40\%). However, due to the large uncertainties, it is not possible to draw firm conclusions on the observed wide range of gas fractions, and more data are needed to set tighter constraints on this property.

It is interesting to compare the amount of gas inferred from the comparison between dynamical and
stellar mass with the amount of gas that can be inferred from the Schmidt-Kennicut law.
More specifically, if we assume the validity of the Schmidt-Kennicut law (hereafter SK) for our $z\sim3$ galaxies, we can
calculate the gas surface density from the star formation rate surface density, and therefore derive the gas mass. In particular we use the relation presented in \cite{Mannucci:2009} (eqs.~1 and 2) that adopt for the slope of the SK relation the value $1.4$ given by \cite{Kennicutt:1998}. To infer
the SFR surface density we use the characteristic radius of the mass distribution $r_e$ estimated in our dynamical modeling as the radius
of the galaxy. Then, under the assumption of a negligible contribution in mass of the dark matter, and by adding the stellar mass to the gas mass we obtain the baryonic mass of the galaxy
($M_{bar}$). In this way we obtain for our objects an estimate of the mass by using only the results of the SED fitting.

\begin{figure}[!ht]
  \centering
  \includegraphics[width=0.99\linewidth]{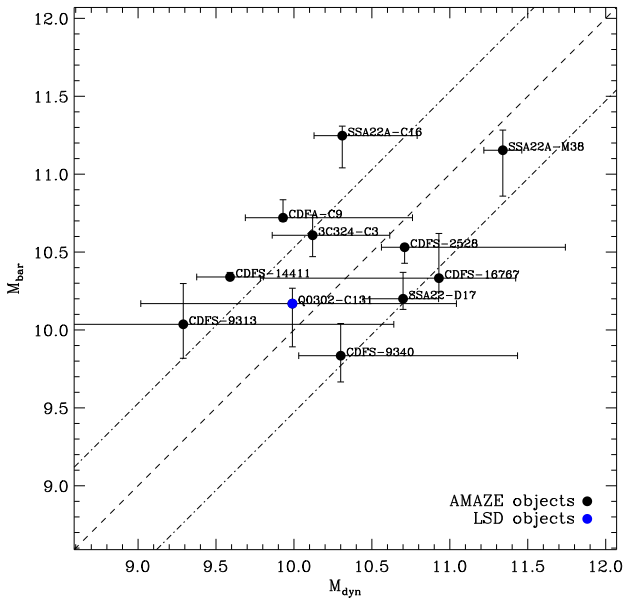}
  \caption{Comparison of the dynamical mass $M_{dyn}$ obtained from the gas dynamical modeling and
  the baryonic mass $M_{bar}$ obtained by combining the stellar mass and the gas
  mass inferred from the SK law. The dashed line represents the condition of equality of the two
  masses. The dot-dashed lines represents the mean scatter of the points around the $M_{dyn}=M_{bar}$ condition ($\sim0.5$ dex).}
  \label{fig15}
  \end{figure}

  In Fig.~\ref{fig15} we show the comparison of the $M_{dyn}$ estimated from the gas dynamics modeling presented in this paper and the $M_{bar}$ estimated from the SK law, combined with the stellar mass. The two estimates of the mass, considering the error bars, are generally quite consistent. The mean difference (in module) (i.e. the mean scatter of the points around the $M_{dyn}=M_{bar}$ condition) is of $\sim0.5$ dex and there is no particular bias of one of the two values on respect of the other. The mean
  residual of the points on respect of the $M_{dyn}=M_{bar}$ condition (i.e. the difference in module divided by the errors) is $\sim1.2$. We can conclude that the results of the SED fitting give a
  reliable estimate of the dynamical mass. This also implies that the two assumptions of the validity of the SK law and of a negligible contribution in mass of the dark matter appear to be valid.

\cite{Daddi:2010a} using CO observations of galaxy samples at low and high redshifts ($z\sim0.5$ and $z\sim1.5$) find a $1.42$ slope for the SK relation, very similar to that by \cite{Kennicutt:1998}. \cite{Genzel:2010} using systematic data sets of CO molecular line emission in $z\sim1-3$ normal star-forming galaxies find instead an index of $\sim1.1-1.2$. In the present work we have used the \cite{Kennicutt:1998} index for comparison with other data in the literature, but using the smaller \cite{Genzel:2010} index we would have obtained, on average, gas masses larger by only $\sim0.3$~dex.
  
\subsection{The Tully-Fisher relation at $z\sim 3$}\label{s53}

The Tully-Fisher relation (hereafter TFR) is the relation between luminosity (or stellar mass) and maximum rotational velocity of disk galaxies \citep{Tully:1977}. This relation, originally used as a distance indicator for disk galaxies, represents an important feature for understanding galaxy formation and evolution because it links directly the angular momentum of the dark matter halo with the stellar mass content of disk galaxies.
In fact, according to theoretical models, galaxy disks form out of gas cooling down from a surrounding
dark matter halo, maintaining its specific angular momentum and settling in a rotationally supported
disk. Therefore, the structure and dynamics of disk galaxies at different cosmic epochs are expected to be closely correlated with the properties of the dark matter halo in which they are embedded.

In recent years, an increasing number of dynamical studies of disk galaxies in the local and
intermediate-redshift Universe allowed the investigation of the evolution of the TFR with redshift.

The evolution of the TFR is expected to be related both to the conversion of gas into stars and to the
inside-out growth of the dark matter halo by accretion. In fact, while the extent of the halo grows
considerably with time, the circular velocity of the halo grows less, keeping the rotation curve
approximately flat to larger and larger distances. The accretion of the dark matter is followed by
accretion of baryonic gas, which is subsequently converted into stars by ordinary star formation in the
disk. However, the details of the process and the amount of evolution expected depends strongly on the
model assumptions, on the accretion mechanism and the timescale needed to convert the gas into stars. Any successful model of disk formation should then be able to reproduce the slope, zero-point, scatter, and redshift evolution of the TFR. 

There are still discrepant results on a possible evolution at intermediate redshifts of the tight
relation observed in local galaxies (e.g., \citealt{Haynes:1999}; \citealt{Pizagno:2007}). For example
\cite{Vogt:1996} reported very little evolution of the B-band TFR up to $z\sim1$, but other groups (e.g.
\citealt{Barden:2003} and \citealt{Nakamura:2006}) found a strong brightening of $\sim1-2$ mag in B-band luminosity over the same redshift range.

Furthermore, the interpretation of a possible evolution of the luminosity TFR is difficult since the
luminosity and the angular momentum might be evolving together. As a consequence, the stellar mass TFR, which correlates the stellar mass and the maximum rotational velocities of disks, offers a physically more robust comparison as it involves more fundamental quantities.

In this context, \cite{Flores:2006}, \cite{Kassin:2007}, and \cite{Conselice:2005} have found no evolution in both the slope and zero point of the stellar mass TFR up to $z=1.2$.

In contrast \cite{Puech:2008} detect an evolution of the TFR zero point of $0.36~dex$ between $z
\sim0.6$ and $z = 0$, by using a sample of 18 disk like galaxies observed with the integral field spectrograph GIRAFFE at the VLT.

At z$\sim$2 \cite{Cresci:2009} build a TFR at $z\sim2$ based on the SINS sample and find an evolution of
the TFR zero point of $\sim0.4~dex$ relative to the local TFR.

 \begin{figure}
  \centering
  \includegraphics[width=0.99\linewidth]{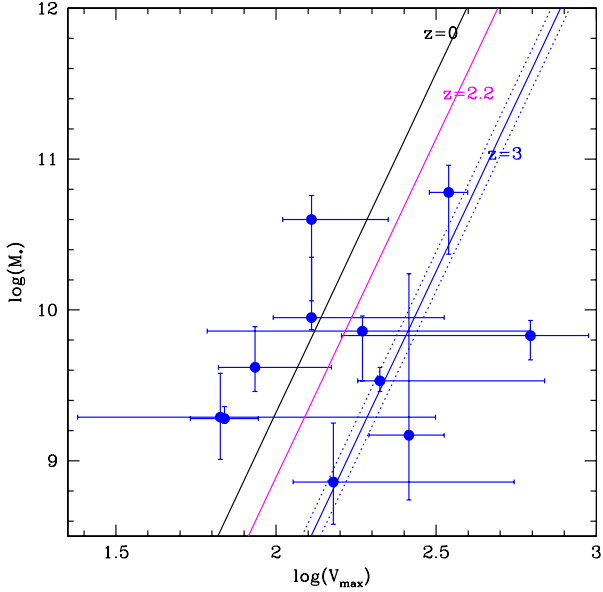}
  \caption{The Tully-Fisher relation at $z\sim 3$ reconstructed by using the dynamical and
  stellar masses of the rotating galaxies found in our sample. The solid black line represents
  the local relation, the magenta line represents the $z\sim2$ relation \citep{Cresci:2009} and
  blue solid line represents the fit to our data ($z\sim3$) by keeping the same slope as for the local and $z\sim2$ cases. The dotted blue lines represents the formal $1\sigma$ uncertainty on the zero point of our fitted relation.}
  \label{fig16}
 \end{figure}

By using the dynamics of $z\sim 3$ galaxies inferred from our data, we are now able to derive the
TFR at $z\sim 3$.
In Fig.~\ref{fig16} we present the $z\sim 3$ TFR obtained by using the data presented in this paper
(obviously only for the rotating objects).
The maximum rotational velocities have been computed from the best fit models as reported
in Tab.~\ref{tab3}. The stellar masses are those reported in table \ref{tab4}.

All our data points are consistent with having a stellar-to-dynamical mass ratio smaller than the local value. We can actually observe an evolution of the TFR with the redshift and this evolution is in the direction predicted by the models: the dynamical mass is already in place at this cosmic epoch, but the stellar mass has yet to be formed.

The large scatter ($\sim$1.5 dex) of the TFR at $z\sim3$ suggests that at this redshift
the relation is not yet in place, probably due to the young age of the galaxies at this epoch  (between $50Myr$ and $500Myr$).
However we can not exclude that our $V_{max}$ measurements are affected by an incorrect assumption
of a rotating disk kinematics for some objects of our samples, which introduces a large scatter in
the relation.

Assuming that the TFR is in place, despite the large data scatter, and that the slope is the same of the
local universe, we can fit the relation obtaining a zero point for our data of $ -0.97\pm0.13$.
This implies an evolution of the zero point of $1.29$ dex relative to the local TFR and $0.88$ dex 
relative to the $z\sim2$ TFR.

 \begin{figure}
  \centering
  \includegraphics[width=0.99\linewidth]{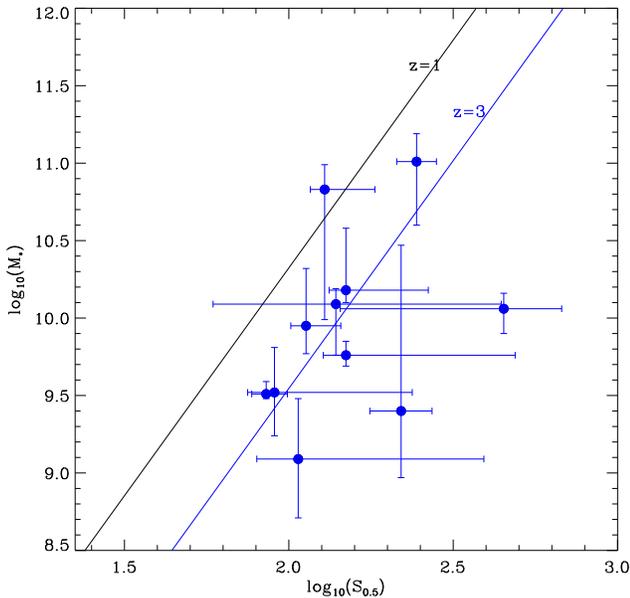}
  \caption{The $S_{0.5}$ Tully-Fisher relation at $z\sim 3$. The solid black line represents
  the $z\sim1$ relation \citep{Kassin:2007}, the dashed line represents the fit to our data ($z\sim3$) by keeping the same slope as for the $z\sim1$ relation.}
  \label{fig17}
 \end{figure}

Given the high intrinsic velocity dispersion observed in our $z\sim3$ sample, we also explored the $S_{0.5}= \sqrt{0.5V^2 +\sigma^2}$ estimator introduced by  \cite{Kassin:2007} to study the impact of turbulent motions. In Fig.~\ref{fig17} we present the $S_{0.5}$ TFR obtained by using our $V_{max}$ and $\sigma_{int}$ values. \cite{Kassin:2007} found a tight relation ($\sim0.3$~dex scatter in $M_{\star}$) in a sample of galaxies with redshift up to $z = 1.2$, with no significant evolution. By only allowing for a change in zero point of the $z\sim1$ relation we estimate a mean scatter of $\sim 0.5$~dex in our galaxies at $z\sim3.3$. The reduction of the scatter in the $S_{0.5}$ TFR compared to the classical one ($\sim1.5$~dex) suggests that, at least for some objects, part of the mass might be supported by turbulent or not ordered motions. We observe a significant evolution of the zero point of the relation between $z\sim1$ and $z \sim3$ ($\sim0.8\pm 0.1$~dex in $M_{\star}$), providing further support for the observed evolution on the classical TFR.

The use of SINFONI integral field spectra represents a major step forward in the modeling of the dynamics of high-z galaxies, as the two-dimensional mapping is more complete and not biased by a priori assumptions about the kinematic major axis and inclination of the system as in long-slit spectroscopy. Also SINFONI data can allow a better selection of the rotating systems and ensure a much lower contamination from complex dynamics and irregular motions than in longslit data sets. This might be a reason for the apparent inconsistency of the results of kinematical studies at redshift greater than $0.5$.

\section{Conclusions}

We conducted an extensive investigation of the dynamics of $z\sim3$ galaxies using near infrared integral field spectra obtained with SINFONI at the ESO VLT. This is the first time that such dynamical analysis is conducted for a relatively large sample (33 galaxies) at $z\sim3$. 

\begin{itemize}
\item We estimated galaxy masses by fitting rotating disk models to selected galaxies. Due to the relatively low S/N of the data, we implemented  a simple but quantitative method to identify smooth velocity gradients, which are the signature of possible disk rotation. After this pre-selection, we performed complete dynamical modeling using a rotating disk model with an exponential mass profile.
\item We found that the kinematics of $\sim 30\%$ of the objects (11 out of 33) is consistent with that of a rotating disk. This fraction is fully consistent with that found with $z\sim2$ samples, like SINS,  suggesting that the fraction of rotating objects does not evolve between $z\sim2$ and 3. 
\item When considering only the galaxies observed with AO and finer pixel scale (LSD sample), only 1 out of 9 shows a smooth velocity gradient. This is likely the consequence of such observational setup providing higher spatial resolution but lower sensitivity to  the outer, low surface brightness regions
which dominate the ``rotation signal''.
\item The 11 rotating objects have dynamical masses in the range $\sim 1.8 \times 10^9 M_{\sun} - 2.2 \times 10^{11} M_{\sun}$ with a mean value of $\sim 1.6 \times 10^{10} M_{\sun}$.
\item By comparing the amplitude of  the inclination-corrected velocity gradient with the intrinsic gas velocity dispersion derived from models, we found that the majority of the rotating galaxies at $z\sim3$ are dynamically ``hot'' disks with an average $\langle\Delta v_{deproj}/\sigma_{int}\rangle \sim 0.9$. This confirms the results obtained by other studies at $z\sim1-2$ ($\Delta v/\sigma \sim 2-4$), indicating  that high-z disk galaxies are more turbulent than local galaxies and even more ``hot'' than galaxies at $z\sim2$ .
\item We compared stellar and dynamical masses for the rotating objects, and derived an estimate of the ``gas and dark matter'' fraction. Using independent estimates for stellar mass and star formation rate, and applying the Schmidt-Kennicut law, we obtained a SED-based estimate of the total baryonic mass of our objects. This estimate is consistent with our estimate of the mass from gas dynamics (rms scatter lower than $0.5$dex). This shows that SED fitting gives a reliable estimate of the dynamical mass and also implies that our assumptions on the validity of the SK law and of a negligible contribution of dark matter are well justified.
\item Finally, we obtained the Tully-Fisher relation at $z\sim3$. All our data points are consistent with a stellar-to-dynamical mass ratio smaller than the value in the local universe confirming the redshift evolution of the relation already found at $z\sim 2$. This is consistent with models predictions, according to which stellar mass is still building up compared to the total dynamical mass. However,  the large scatter of the points may also suggest that, at this redshift, the relation is not yet in place, probably due to the young age of the galaxies.
The Tully-Fisher relation based on the $S_{0.5}$ indicator has a significantly smaller scatter that the ``classical'' one and also shows an evolution in zero point compared to $z\sim1$. The reduction in scatter suggests that, at least for some objects, part of the mass might be supported by turbulent motions.

 \end{itemize}

\begin{acknowledgements}
We wish to thank the referee for his/her really thorough and constructive report on this paper. This work was partly funded by INAF and ASI through contract ASI-INAF I/016/07/0 and by the Marie Curie Initial Training Network ELIXIR of the European Commission under contract PITN- GA-2008-214227. 
\end{acknowledgements}

\bibliographystyle{aa} 

\end{document}